\documentclass[twocolumn,superscriptaddress,prb,aps,amsmath]{revtex4-1}

\usepackage{graphicx}

\begin{document}

\title{Group-velocity slowdown in a double quantum dot molecule}

\author{Stephan Michael}
\affiliation{Department of Physics and Research Center OPTIMAS, University of Kaiserslautern, P.O. Box 3049, 67653 Kaiserslautern, Germany}
\author{Weng W. Chow}
\affiliation{Semiconductor Materials and Device Sciences Department, Sandia
  National Laboratories, Albuquerque, NM 87185-1086, USA} 
\author{Hans Christian Schneider}
\affiliation{Department of Physics and Research Center OPTIMAS, University of Kaiserslautern, P.O. Box 3049, 67653 Kaiserslautern, Germany}

\begin{abstract}
The slowdown of optical pulses due to quantum-coherence effects is
investigated theoretically for an ``active material'' consisting of
InGaAs-based double quantum-dot molecules. These are designed to exhibit a long lived
coherence between two electronic levels, which is an essential part of a
quantum coherence scheme that makes use of electromagnetically-induced transparency effects to achieve group velocity slowdown. We apply a many-particle approach based on realistic
semiconductor parameters that allows us to calculate the quantum-dot material
dynamics including microscopic carrier scattering and polarization dephasing
dynamics. The group-velocity reduction is characterized in the frequency
domain by a quasi-equilibrium slow-down factor and in the time domain by the
probe-pulse slowdown obtained from a calculation of the spatio-temporal
material dynamics coupled to the propagating optical field. The group-velocity
slowdown in the quantum-dot molecule is shown to be substantially higher than
what is achievable from similar transitions in typical InGaAs-based single
quantum dots. The dependences of slowdown and shape of the propagating probe pulses on lattice temperature and drive intensities are investigated. 
\end{abstract}

\maketitle

\section{Introduction}

Quantum coherence effects encompass a variety of interference effects in the
coherences, i.e., transition amplitudes, between quantum states that are driven by laser light. In quantum optics, they have
been known for decades.~\cite{Fleischhauer,intro4,scully1,intro2,intro3,mompart} 
In particular, electromagnetically induced transparency (EIT) is based on
the quantum interference associated with a long-lived coherence, which can make an optically thick medium transparent for a
probe field in the presence of a drive field. Because the coherent
effects also modify the dispersive properties, a very small group
velocity may occur for pulses, which is usually referred to as slow light. 
Electromagnetically induced transparency, group velocity slowdown and other quantum-coherence effects have been intensively investigated
in atomic, molecular and optical (AMO) physics, see, e.g.,
Refs.~\onlinecite{lukin:nature05:stationary-light,Fleischhauer}. 
There have been different proposals to realize quantum coherence effects
in few-level systems in solid state~\cite{intro10,intro7,yang1,sarkar1,chuang-prb04:exc-pop-pulsation,boyd-pra04:coupled-resonator,Nikonov1,intro5} and especially
semiconductors,~\cite{intro8,intro9,intro11,hau1,intro6,intro13,intro10,yang1,sarkar1}
because of the possible importance of these effects for optical
information processing, such as an optical delay line. Slow light has been
achieved in semiconductor quantum wells using setups that employ coherent
population oscillations of excitons instead of the EIT-type processes in
quantum dots (QDs) considered in the present paper.~\cite{chuang-prb04:exc-pop-pulsation,Palinginis-APL} Other approaches, for instance, involving slow light in photonic crystals, are also being actively pursued.~\cite{Kondo}

Semiconductor QDs, which are arguably the closest realization of a
system with localized states and discrete energies in semiconductors, are a
natural candidate for the realization of quantum-coherence
processes~\cite{chang1,intro12,qcpinsqd,Nielsen2,mork_JOSAB} in a material,
for which extremely advanced growth and processing techniques exist. However,
for electron-hole transitions in semiconductors typical dephasing
times severely limit the achievable group velocity
slowdown, even in QDs,~\cite{mork_JOSAB,NielsenProp} where there is the
smallest ``phase space'' for scattering and dephasing processes. 
As it is known from quantum optics, such a pronounced dephasing is detrimental
for quantum coherence effects. Depending on
the levels that are connected by drive and probe fields,  $\Lambda$, $V$ and
ladder schemes can be realized, and these can be compared directly as long as
one applies an AMO model that assumes dephasing constants for the
various polarizations involved in the respective schemes.~\cite{mork_JOSAB}
For instance, it has been shown that the structural QD parameters can generally be more
easily optimized for $V$ schemes than for other
schemes.~\cite{Nielsen1,Nielsen2} 

We will discuss in this paper only $V$-type schemes, as opposed to the $\Lambda$ schemes analyzed earlier by us,~\cite{qcpinsqd,apl:quantum-coherence,jmo}
in which the quantum coherence
connected two hole states. In such a $\Lambda$ setup there is a sizable dephasing of the quantum coherence from the hole states because
they are closely spaced and broadened by polaronic interaction effects. This problem can partly be circumvented by using a short
drive pulse,~\cite{apl:quantum-coherence,jmo} but the time window, during which the
probe pulse is slowed down, is too short to be useful for applications.~\cite{dissertation} 

In this paper we make a theoretical proposal for a QD
\emph{molecule} that is designed to lead to a long-lived coherence between its lowest electronic levels. The proposed design consists
of two QDs of different sizes stacked in growth direction, and should be
within reach of current fabrication techniques, as evidenced by recent
investigations that have shown how QD molecules can be fabricated with
prescribed properties.~\cite{QDH2,Kapon,Coleman}  The line-up of the energy levels
of the proposed QD molecule are not qualitatively different from those of
a single InGaAs-based QD, for which quantum-coherence effects have already been
investigated.~\cite{qcpinsqd,apl:quantum-coherence,jmo,Nielsen1} What sets
the molecule apart from the single QD is the ``wave-function engineering''
that leads to dipole matrix elements and dephasing rates that are favorable
for group-velocity slowdown. We demonstrate this by employing a model from
semiconductor many-particle physics rather than an AMO model. We
stress that, while an AMO model uses constant dephasing rates for the
individual levels and the dipole matrix elements as input, our approach uses
the relevant matrix elements from a simplified QD electronic structure calculation
and computes, in a microscopic fashion, the dephasing and scattering processes
in QD molecules due to the Coulomb interaction between charged carriers and/or
the carrier-phonon interaction. Based on this approach, we characterize the
slow-down factor of these QD molecules in the frequency domain and explicitly calculate
the group-velocity slowdown of a probe pulse propagating in a semiconductor
host surrounding these model QD molecules. For the dynamical calculation, we
combine a determination of the propagating optical fields, as in
Refs.~\onlinecite{mork_JOSAB,NielsenProp}, with a microscopic theory for the
calculation of scattering and dephasing processes along the lines of
Refs.~\onlinecite{prb70:235308,Jahnke1,Jahnke2,Jahnke3,Jahnke4,Jahnke5}. Work in this area has recently been comprehensively reviewed in Ref.~\onlinecite{review_chow_jahnke}. 

The paper is organized as follows: The design of the asymmetric QD molecule is presented along with a calculation of its
electronic single-particle states and energies in
Sec.~\ref{sec:model}. Results from the semiconductor Bloch equations
including many-particle scattering and dephasing effects as they apply
to QDs and from the treatment of pulse propagation in
semiconductors are gathered in Sec.~\ref{sec:semiconductor-MBE}.
Using the semiconductor Bloch equations, we first
investigate the slowdown factor and the slowdown-bandwidth product
determined from the spectral features for different lattice
temperatures and cw drive intensities in
Sec.~\ref{subsec:spectra}. Next, in
Sec.~\ref{subsec:pulse-propagation} we determine the slowdown factor
and pulse characteristics directly from the propagating probe
field. We investigate the influence of the probe-pulse shapes for
different lattice temperatures, and cw drive intensities.
We compare these results with the group velocity slowdown determined
from the spectral features. The final discussion of
Sec.~\ref{sec:results_single} concerns the differences to quantum-coherence
schemes in single semiconductor QDs. The setup for the single QDs including
electronic single-particle states and energies is presented in 
Sec.~\ref{subsec:results_single_model}. The slowdown factor determined from the
spectral features for different lattice temperatures and
cw drive intensities is investigated in Sec.~\ref{subsec:results_V_single}. These results are compared to the QD
molecule results from Sec.~\ref{subsec:spectra} in Sec.~\ref{subsec:results_single_comparison}.      
We present our conclusions in Sec.~\ref{sec:conclusion}.


\begin{figure}[tb]
\centering
\includegraphics[trim=5cm 2cm 4cm 2cm,clip,scale=0.5,angle=0]{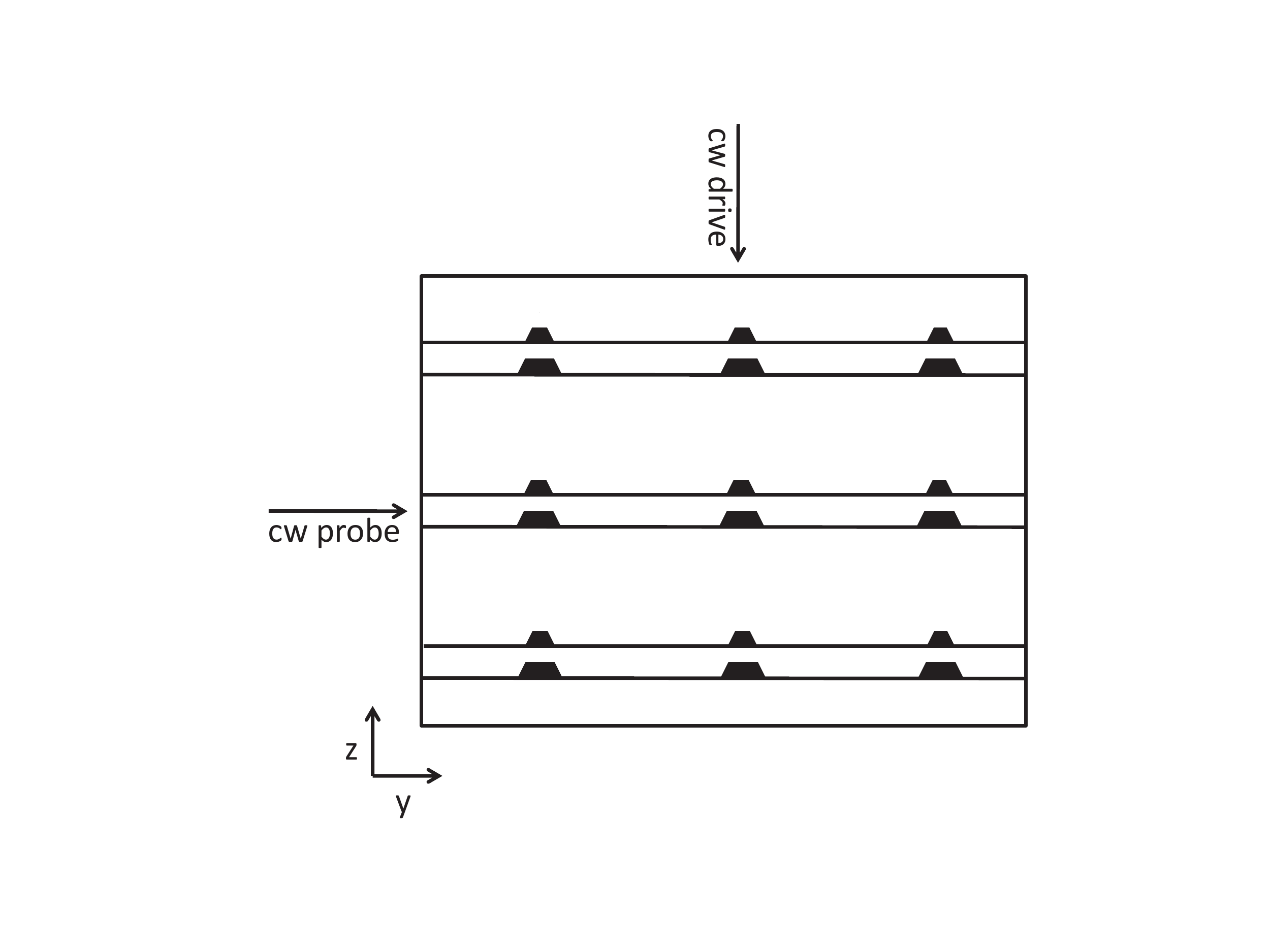}
\caption{Schematic drawing of setup for the calculation of group velocity slowdown with a V-scheme formed by levels in the QD molecules as shown in Fig.~\protect\ref{figure1b}. For the determination of the slowdown factor, the drive field is assumed to be cw; for the dynamical calculation a ``quasi-cw'' drive pulse with duration much larger than the probe pulse is used.}
\label{figure1a}
\end{figure}

\section{Electronic Structure of the QD Molecule \label{sec:model}}

The purpose of this section is to introduce the design of an asymmetric QD molecule with an electronic structure that is particularly well suited for quantum coherence effects in a $V$ configuration. We do not aim at a comprehensive theory of the electronic structure of QD molecules, which would have to include the complicated
alloy concentration and strain fields of the QD molecule and the surrounding structure. Rather, we describe here a numerically tractable model that 
includes the geometry of the QD \emph{molecules} under study as well as static
electric fields and works with a few meaningful parameters that characterize
the structure. In particular, we compute the single-particle states, i. e., wave functions
and energies of the double QD molecules from the states of the two
underlying single QDs that make up the molecule.  In the following, we assume
that the QD molecules are formed from two vertically stacked QDs, separated by
a spacer layer, such as one of the three double layers sketched in
Fig.~\ref{figure1a}. We assume that the QD molecules are embedded in a quantum
well, and take the QW continuum states as plane waves that are orthogonalized
to the localized states of the QD molecule, see Ref.~\onlinecite{prb64:115315}.

For the QDs we assume that they are grown on a wetting
layer embedded in a quantum well. We use the envelope-function approximation and assume a cylindrical
confinement potential of finite depth which yields semi-analytical
results for the wave functions and energies. This approach is described in detail in appendix~\ref{app_dot}. We stress that, although no strain,
piezoelectric effects or structural anisotropies are included in the single QD model, its parameters are adjusted to more accurate $k\cdot p$ QD calculations~\cite{homepage,hackenbuchner} with material parameters as used in Ref.~\onlinecite{jmo}.   

We start with the description of the two separate QDs that will be combined to the double-QD molecule. Since the QD molecule should be asymmetric, we identify the single QDs as the ``small'' and the ``large'' one.  For the small QD, we assume an In$_{0.8}$Ga$_{0.2}$As QD embedded in a GaAs quantum well on a wetting
layer of thickness 1\,nm. The QD has a diameter of 10\,nm and a height of
2\,nm. For this QD geometry only the lowest electron and hole states
are confined. For the large QD, we assume an In$_{0.9}$Ga$_{0.1}$As QD embedded in
a GaAs quantum well on a wetting layer of thickness 1\,nm. The
cylindrical QD model has a diameter of 12\,nm and a height of
3\,nm. For this structure three electron and three hole states are
confined. The different sized QDs have different energy spacings between the levels. In particular, no energetic degeneracies between levels of the small
and the large QD occurs.

\begin{table}[bt]
\begin{tabular}{||l|c||l|c||}
\hline\hline
state & $E_e$ (meV)&state& $E_h$(meV) \\ 
\hline
e$_{0}$ & $-194$ & h$_{0}^{\text{b}}$ & $55$  \\
e$_{1}$ & $-132$ & h$_{1}^{\text{a}}$ & $47$\\ 
e$_{2/3}$ & $-56$ & h$_{2/3}$ & $13$ \\
\hline\hline
\end{tabular}
\caption{Electron (e) and hole (h) energies of single-particle states
  in the QD molecules. The bonding and antibonding states formed from hole levels of the individual QDs are denoted by h$^{\text{b}}$ and h$^{\text{a}}$, respectively.}
\label{table-1}
\end{table}

\begin{figure}[tb]
\centering
\includegraphics[trim=6cm 2cm 4.5cm 3cm,clip,scale=0.5,angle=0]{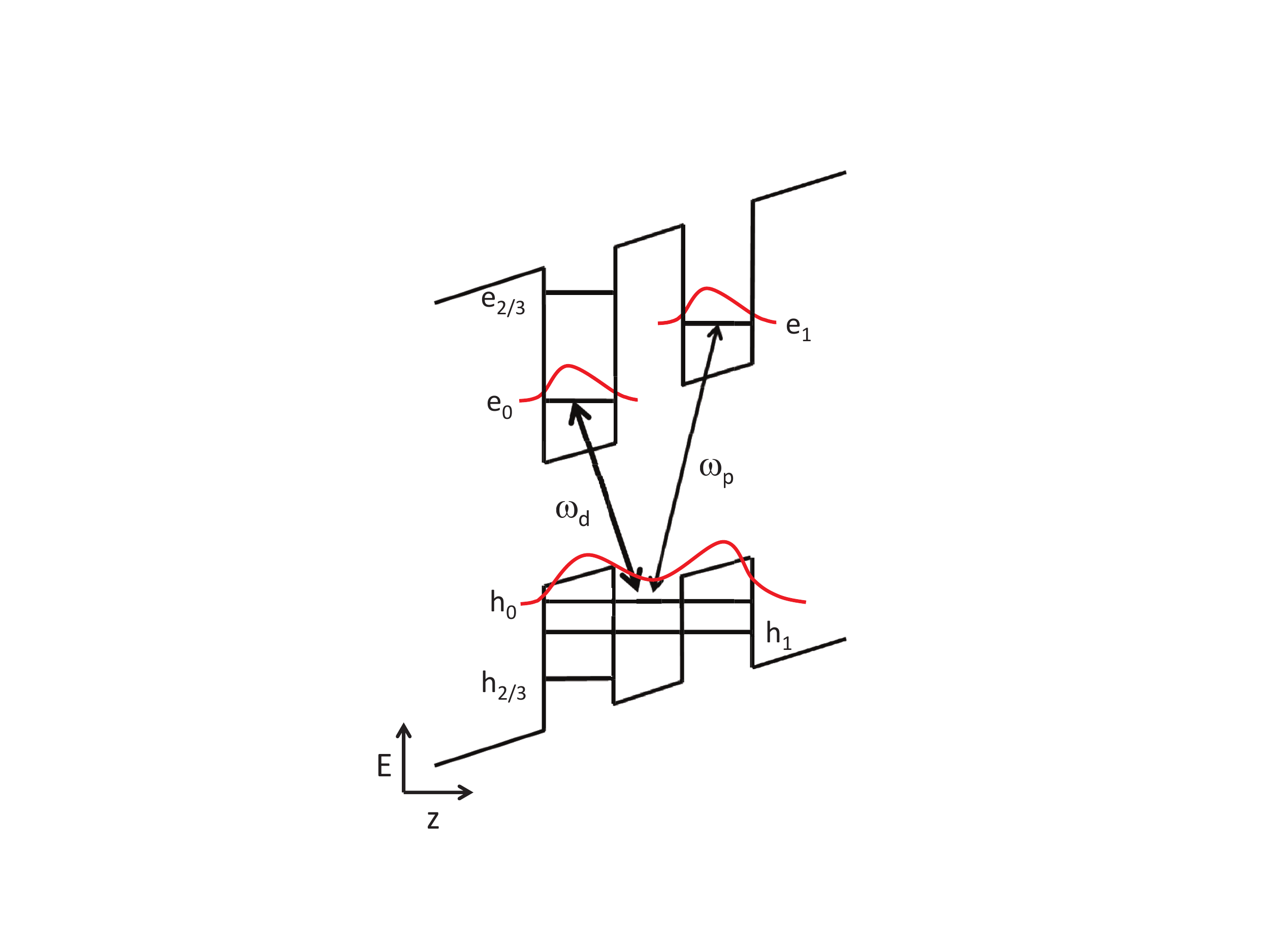} 
\caption{Schematic picture of the lowest-level wave functions and the geometry of the asymmetric double-QD molecule described in the text.
The resonant probe and drive fields in a $V$-type quantum coherence scheme are also shown.}
\label{figure1b}
\end{figure}

The states of the two cylindrical QDs described above are the limiting case
for a level structure of a QD molecule composed of the individual QDs, but
with a very large spatial separation between the two.  When the QDs are closer
together with a potential barrier between them, the levels of the original 
QDs are mixed to form the single-particle states of the QD molecule. The
states of the QD molecule are obtained from a linear combination of the
orbitals of the QDs, as described in Appendix~\ref{app_molecule}. In this
approach, the effect of external static electric fields is also included. The
QD molecule is designed by choosing a combination of an external field in growth direction and a separation of the QDs, so that the lowest hole levels of the two QDs are lined up without bringing the QDs too close to together. We choose a static electric field in growth direction of $E_{\perp}=1.5$\,mV/nm and a QD distance of 14\,nm, placing the QD molecule
in the center of a 30\,nm surrounding quantum well. For the QD molecule we obtain four confined hole and electron
states whose energies are compiled in Table~\ref{table-1}. The line-up of the levels is shown schematically together with a sketch of the most important wavefunctions in Fig.~\ref{figure1b}. There is only a very small overlap between wave functions of the lowest electronic levels e$_{0/1}$, which are only very weakly mixed states that are mainly localized in the individual QDs. The lowest QD hole levels, however, are bonding h$_{0}^{\text{b}}$ and antibonding h$_{1}^{\text{a}}$ states formed from the lowest hole levels in the individual QDs. Further, the transitions
between the lowest bonding hole level h$_{0}^{\text{b}}$ and the electron levels e$_{0}$ and e$_{1}$ are dipole allowed with dipole moments of $0.5e$\,nm and $0.2e$\,nm, respectively.

The design of the QD molecule thus leads to a level structure and dipole
moments that are especially well suited for a $V$-configuration with probe and
drive fields connecting the h$_{0}^{\text{b}}\leftrightarrow\text{e}_{1}$ and
h$_{0}^{\text{b}}\leftrightarrow\text{e}_{0}$ states as shown in
Fig.~\ref{figure1b}.  Quantum-coherence effects in the $V$ scheme are
particularly pronounced if the transition e$_{0}\leftrightarrow\text{e}_{1}$
is long lived, i.e., exhibits only a small polarization dephasing. This is the transition that is ``engineered'' to
connect electronic states of the QD molecule that are more or less localized
in the different individual QDs and therefore have a very small wave-function
overlap. We refer to the accompanying polarization as the \emph{quantum coherence}.


\section{Semiconductor Maxwell-Bloch equations\label{sec:semiconductor-MBE}}

In this section, we summarize the equations for the propagating
optical fields and the coupling to the semiconductor Bloch equations, which
are used to describe the $V$-scheme of the QD system.
The propagating optical field is written in the form
\begin{equation}
\vec{E}( y,t) =\frac{1}{2}\hat{x}[ \mathcal{E}(y,t)e^{i( ky-\omega t)}
    +\mathcal{E}(y,t) e^{-i( ky-\omega t) }] \label{E1}
\end{equation}
where the propagation is in $y$-direction, $\hat{x}$ is the polarization unit vector in $x$ direction, 
$k$ is the wavevector and $\omega$ is the frequency of the field $\vec{E}$. The corresponding macroscopic
polarization has the form
\begin{equation}
\vec{P}(y,t) =\frac{1}{2}\hat{x}[ \mathcal{P}(y,t) e^{i(ky-\omega t) }
    +\mathcal{P}^{*}(y,t)e^{-i( ky-\omega t) }]  \label{P1}
\end{equation}
where $\mathcal{P}$ is the complex slowly varying envelope. Substituting these
forms into the wave equation and employing the slowly-varying envelope
approximation, one obtains the slowly-varying Maxwell equations.~\cite{oldprop1,oldprop2} The
substitution $t'=t-n_{b} y/c$, where $n_{b}$ is the background refractive index, transforms the partial
differential equation into an ordinary differential equation with respect to
the scaled spatial variable $y$
\begin{equation}
\frac{\partial}{\partial y}\mathcal{E}
    =i\frac{n_{b}\omega }{2\varepsilon c}\mathcal{P}
\label{svmaxwell}
\end{equation}
This equation neglects the dependence on the transverse and lateral coordinate of the fields. We assume in the following that the lateral and transverse extension of the probe pulse, which is propagated by Eq.~\eqref{svmaxwell} in our setup, see Fig.~\ref{figure1a}, is such that this approximation is fulfilled.

The macroscopic polarization $P$ is connected with the microscopic
polarization by
\begin{equation}
P=\frac{N_{d}}{L}\sum_{\alpha ,\beta }\mu _{\alpha\beta }p_{\alpha \beta } + c.c.
\label{eq_makro}
\end{equation}
where $N_{d}$ is the density of QDs in the quantum well layer, $L$ is
the thickness of the quantum well, in which the QDs are
embedded, $\mu_{\alpha\beta }$ are the dipole matrix elements and
the summation index $\alpha$ or $\beta$ refers to QD system
electron or hole states, respectively.


\subsection{Semiconductor Bloch equations}

The dynamics of the polarizations and carrier distributions at the
single-particle level are calculated in the framework of the
semiconductor Bloch equations for the reduced single-particle density
matrix. We denote in the following electron and hole levels in the QD $\alpha$ and~$\beta$, respectively.
For the
$V$ system of interest in this paper one obtains the following
equations of motion for the ``interband'' polarizations,
$p_{\alpha \beta}$, and the ``intra(electron-)band'' polarizations~$p_{\alpha'\alpha''}$
\begin{align}
\begin{split}
\frac{\partial}{\partial t}p_{\beta\alpha } =&  - i \omega_{\alpha \beta }p_{\beta \alpha } - i\Omega _{\alpha \beta }\left(n_{\alpha }^{c}-n_{\beta }^{v}\right) \\ 
&  -i\sum_{\alpha'\neq \alpha}\Omega _{\alpha'\beta}p_{\alpha'\alpha}+S_{\beta \alpha}
\label{p-ab}
\end{split}\\
\begin{split}
\frac{\partial }{\partial t}p_{\alpha'\alpha''}
=& - i\omega _{\alpha''\alpha'}p_{\alpha'\alpha''}-i\Omega_{\alpha''\alpha'}\left( n_{\alpha''}^{c}-n_{\alpha'}^{c}\right) \\ 
&   +i\sum_{\beta'}\left( \Omega_{\alpha''\beta'}p_{\alpha'\beta'}-\Omega_{\beta'\alpha'}p_{\beta'\alpha''}\right) +S_{\alpha'\alpha''}
\label{p-aa}
\end{split}
\end{align}
In particular, the polarization $p_{e_{0} e_{1}}$ here is the \emph{quantum coherence}.
For the time evolution of the electron and hole populations,
$n_{\alpha }^{c}$ and $n_{\beta }^{v}$, one obtains
\begin{align}
\frac{\partial}{\partial t}n_{\alpha}^{c} &=i\sum_{\beta'}
\left( \Omega_{\alpha \beta'}p_{\alpha\beta'}
-\Omega_{\beta'\alpha}p_{\beta'\alpha}\right) +S_{\alpha \alpha } \label{n-a} \\
\frac{\partial }{\partial t}n_{\beta }^{v}& =i\sum_{\alpha'}
\left( \Omega _{\beta \alpha'}p_{\beta\alpha'}-\Omega_{\alpha'\beta}p_{\alpha'\beta}\right) +
S_{\beta \beta }
\label{n-b}
\end{align}
The coherent contributions of the above equations contains transition frequencies $\omega_{\alpha \beta}$ and renormalized Rabi
frequencies $\Omega_{\alpha \beta}=\hbar^{-1} \mu_{\alpha \beta} E \left(
t \right) + \Omega_{\alpha \beta}^{\text{HF}}$. Here, $E(t)$ is the electric field at the position of the QD and the
excitation-dependent Hartree-Fock (HF) contributions~$\Omega_{\text{HF}}$ result from the
Coulomb interaction, as discussed, e.g., in Refs.~\onlinecite{jmo,apl:quantum-coherence,qcpinsqd}. 

The correlation contributions are generally denoted by $S$ and contain the influence of electron-electron and electron-phonon interactions beyond the Hartree-Fock level. In particular,
$S_{\alpha,\alpha}$ and $S_{\beta,\beta}$ describe scattering contributions in the dynamical equations for the electron and hole distributions as well as
dephasing $S_{\beta \alpha}$, $S_{\alpha'\alpha''}$ in the dynamical equations for the coherences. 

\subsection{Scattering/dephasing contributions\label{sec:corcontr}}

Dephasing processes are extremely important for the description of pulse
slowdown in semiconductors. From quantum optics it is well known that the
dephasing rate of the quantum coherence has a decisive influence on the behavior of quantum-coherence schemes. In fact, the engineering of the QD molecule was done with the goal of realizing a comparatively small dephasing rate for the quantum coherence between the $e_{0}$ and $e_{1}$ levels. For short-pulse dynamics, also the population dynamics play a role, and we therefore have to examine both dephasing and scattering contributions to the semiconductor-Bloch equations as they apply to our proposed QD molecule. Scattering processes in the QDs connect discrete levels so that the influence of level broadening is much more pronounced than for scattering between continuum states in quantum wells.

We are here concerned with a treatment of carrier
relaxation and polarization dephasing that captures the essential features for the
analysis of our quantum-coherence scheme. 
To begin with, the broadening of QD levels is mainly provided by the interaction of electrons with phonons and with other electrons in the scattering continuum, which we assume to be formed in the quantum well embedding the QDs. We will only be concerned with excitation conditions in which the continuum states are not appreciably populated by carriers. In this case of vanishing excitation of the continuum states, the electron-phonon interaction has been shown to dominate over the electron-electron interaction for scattering processes and dephasing processes that can be associated with real scattering transitions (as opposed to ``pure dephasing'' processes). We will therefore consistently neglect electron-electron interactions for both of these processes and describe first our treatment of the electron-phonon interaction.

Since the broadening of the discrete levels is important for QDs, it is more appropriate to work with polarons, i.e., quasiparticles that include the effect of the coupling to phonons, instead of the ``naked'' QD electronic levels. Qualitatively, the polaron spectrum contains a peak at the ``naked'' electron energy as well as sidebands due to coupling to the discrete LO phonons. Coupling to a continuum, such as acoustic phonons adds an additional broadening to the peak and the sidebands. In this case, the relaxation and dephasing contributions for the carrier distributions and polarizations cannot easily be computed using Fermi's Golden Rule arguments because there is no straightforward energy conservation for transitions between polarons. Instead, we follow Refs.~\onlinecite{Jahnke2,Jahnke3,Jahnke4} and obtain the scattering and dephasing contributions from the Keldysh Green function technique. In particular, we employ the random-phase approximation (RPA) for the electron-phonon interaction contributions to the electron, or rather, polaron self energy. 

Our treatment of scattering and dephasing contributions is described in Appendix~\ref{app_corcontr}, here we only summarize our approach.
As shown in Ref.~\onlinecite{Jahnke3}, the full polaronic dynamics is, in principle,
not determined by equations of the form~\eqref{p-ab}--\eqref{n-b}, but
rather by coupled equations of motion for ``spectral'' and ``kinetic'' Green
functions \emph{depending on two time arguments} whose numerical solution is
extremely demanding. We therefore follow the spirit of Ref.~\onlinecite{Jahnke4} and separate the
spectral properties of the polarons in order to get equations of motion for
the dynamical distributions and polarizations as defined above. This procedure
yields scattering and dephasing contributions of the form appearing in
Eqs.~\eqref{p-ab}--\eqref{n-b} that still include memory integrals with
information about the polaronic spectrum. 

In contrast to Ref.~\onlinecite{Jahnke4}, we use a Markov approximation and
introduce an \emph{effective quasi-particle broadening} in the memory integrals. This finally yields the scattering and
dephasing contributions as employed in the following calculations.  
The explicit expressions are given in appendix~\ref{app_corcontr} and contain the effect of the electron-phonon interaction on the polaronic spectrum in the form of complex renormalized energies of a single-particle QD state~$\lambda$
\begin{equation}
\tilde{\epsilon}_{\lambda}=\epsilon_{\lambda}+\Delta\epsilon_{\lambda}-i\Gamma_{\lambda}
\label{complexenergy}
\end{equation}
where $\Delta\epsilon$ contains the real Hartree-Fock energy shift and a small
correlation contribution. The broadening~$\Gamma_{\lambda}$ of the level $\lambda$ is entirely due to correlations. We will use in the following a constant level broadening $\Gamma =0.84\,\text{ps}^{-1}$ for a lattice temperature of~$T_{\text{L}}=300$\,K and $\Gamma = 0.50\,\text{ps}^{-1}$ for $T_{\text{L}}=150$\, K,
respectively.~\cite{dissertation} Although the precise value
of $\Gamma$ does not affect the numerical results for our quantum coherence
scheme, it is important to get its order of magnitude right, and we
have determined these numerical values from single-pole approximations to the zero-density QD
polaronic spectral functions, computed as in Refs.~\onlinecite{Jahnke4,Jahnke2,Jahnke-EPJB}.

As mentioned above, we neglect the Coulomb-interaction contribution to
scattering and dephasing processes that involve continuum states. 
However, there are pure-dephasing contributions from the Coulomb interaction between carrier states in the QD, most notably processes in which two electrons effectively exchange their single-particle
states. We therefore take the Coulomb interaction between states \emph{in the QD} into account, because the electron-phonon
contribution to the dephasing of the quantum coherence can be very inefficient, especially for deep QDs, in which the energy conservation between the widely spaced
electronic levels and a LO phonon cannot be fulfilled, even including the
polaronic broadening. As in the case of the carrier-phonon interaction, we
follow Refs.~\onlinecite{Jahnke1,Jahnke4} for the treatment of the Coulomb
interaction contribution and employ the self-energy in second
order Born approximation along with the Markov approximation in the
scattering kernels, as well as a single-pole approximation for the polaronic
spectral properties.
Because the continuum states are not appreciably populated by carriers we
neglect, in contrast to Ref.~\onlinecite{Jahnke4}, the Coulomb-interaction
contribution to the \emph{effective quasi-particle broadening}. 
The relevant equations, including a statically screened
Coulomb potential, have the structure as shown in appendix~\ref{app_corcontr}.


\section{Numerical Results for the QD molecule\label{sec:results}}

In this section we present numerical results with the eventual aim to characterize the slow-down achievable in a structure composed of single layer of double QD molecules for the geometry shown in Fig.~\ref{figure1a}. We assume a strong cw drive field in $z$ direction and a weak cw or probe field in $y$ direction. As we do not include propagation effects for the drive field, our results also apply to multi-layer QD molecule structures, as already indicated in Fig.~\ref{figure1a} as long as the layers are stacked tightly enough in $z$ direction that propagation effects are unimportant and if the propagating field is guided such that it overlaps well with the QD active material. 

Because of the drive field induced energy shifts of the sharp and closely spaced resonances, it is advantageous to first neglect propagation effects for the probe field, and analyze the spectral properties experienced by a weak cw probe field in Sec.~\ref{subsec:spectra}. Although the measure of the achievable slowdown that can be obtained from the spectra is not as accurate as the slowdown for propagating probe pulses computed in Sec.~\ref{subsec:pulse-propagation}, the spectra yield important information on the width of the spectral region in which slowdown is possible and also on the magnitude of the slowdown when studying the influence of parameters, such as drive intensity or temperature. Further, the spectral information is necessary to properly tune the probe pulse in order to maximize the slowdown.

\subsection{Spectra and slowdown factor\label{subsec:spectra}}

\begin{figure}[tbp]
\centering
\includegraphics[trim=6cm 0cm 7cm 0cm,clip,scale=0.65,angle=0]{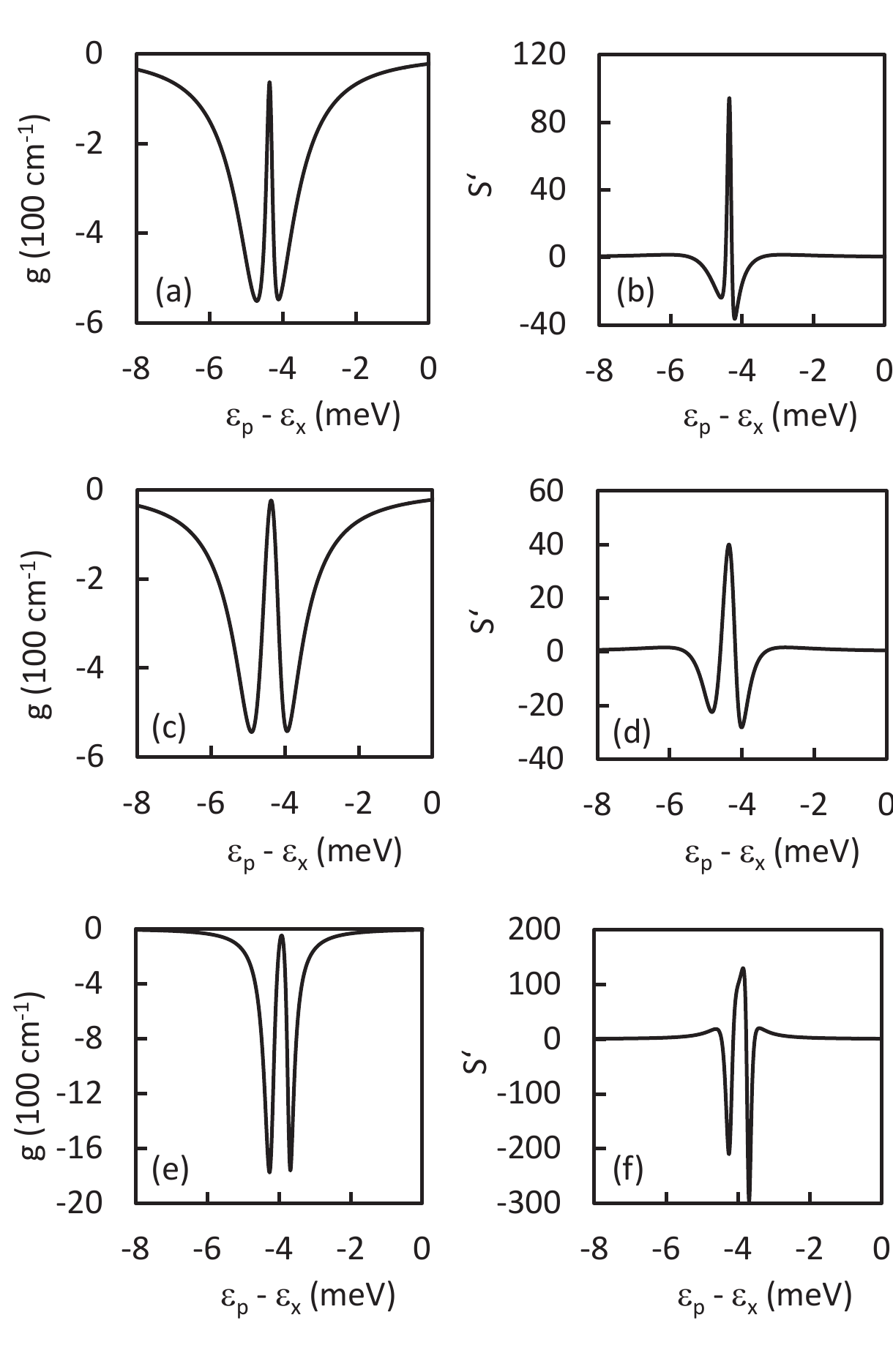}
\caption{Gain $g$ and slowdown $S'$ versus probe detuning for: $T=300$\,K, $I_d= 0.25\,\text{MW/cm}^{2}$ (a,b);  $T=300$\,K, $I_d=0.7\,\text{MW/cm}^{2}$ (c,d) and $T=150$\,K, $I_d=0.25\,\text{MW/cm}^{2}$ (e,f). The unexcited exciton transition energy is denoted by $\epsilon_{x}$.
\label{figure2}}
\end{figure}

We first investigate the spectral features of the slowdown factor and the spectral width over which slowdown can be achieved. To this end, we solve the dynamical equations~\eqref{p-ab}--\eqref{n-b} for a strong cw drive field with fixed angular frequency $\omega_d$ and a weak cw probe field with angular frequency $\omega_p$. From the steady-state value of the polarization $\mathcal{P}$ we determine the gain via
\begin{equation}
g(\omega_{p}) =-\frac{\omega_{p}}{2\varepsilon_{0}n_{b}\mathcal{E}_{p}}
\Im[\mathcal{P}]
\end{equation}
and refractive-index change
\begin{equation}
\delta n(\omega_{p}) =-\frac{1}{2\varepsilon_{0}
  n_{b}\mathcal{E}_{p}}\Re[\mathcal{P}]
\end{equation}
where $n_{b}$ is the background refractive index of the host
material. The group-velocity slowdown factor is defined by
$S\left( \omega _{p}\right) =n_{b}
+\omega_{p} \frac{d\left( \delta n\right) }{d\omega_{p}}\equiv n_b + S'(\omega_p)$, but we will consider only the contribution from the index change 
\begin{equation}
S'(\omega_p)=\omega_{p} \frac{d\left( \delta n\right) }{d\omega_{p}}
\end{equation}
in order to remove the static contribution, which describes the change in group velocity due to the background refractive index as compared to vacuum.
For the numerical calculations we assume a lattice temperature of $T=300$\,K and a cw
probe with a field intensity of $I_p=45\,\text{W/cm}^{2}$. For a cw drive
intensity of $I_d=0.25\,\text{MW/cm}^{2}$ we obtain the spectra shown in
Fig.~\ref{figure2}~(a,b) and for a cw drive intensity of
$I_d=0.7\,\text{MW/cm}^{2}$ the ones shown in Fig.~\ref{figure2}~(c,d). Before
discussing these spectra in some detail, we emphasize that the width of the
spectral features in Fig.~\ref{figure2} is \emph{not} due to effective
dephasing rates for the different polarizations in the system. Instead, the spectral location and the width of the features is entirely due to the calculated dephasing (and scattering) contributions, which are determined by the electronic structure of the QD molecule and the excitation conditions. Nevertheless, one can attempt to extract effective dephasing rates for three-level systems for specified excitation conditions. We defer this question to the end of Sec.~\ref{subsec:pulse-propagation}.


The HF corrections lead to renormalizations of the transition
frequencies as well as of the generalized Rabi frequencies when the excitation, i.e., the drive intensity, is increased. In particular,
excitation dependent HF energy corrections lead to an energy shift of
approximately $4.2$\,meV in Figs.~\ref{figure2} (a)--(d). Also a
small asymmetry, more pronounced for the slowdown factor and less pronounced
for the gain, occurs due to the influence of the HF corrections.

Figure~\ref{figure2} (a)--(d) shows the typical signatures of
EIT~\cite{AutlerTownes,Fleischhauer}: a dip in the absorption profile and an
increase of the slowdown factor $S'$ at the dip. For the sake of simplicity,
we will refer to the existence of two transitions, which are ``dressed'' by
the strong coherent drive field, as Autler-Townes splitting. This splitting is
proportional to the drive intensity. The EIT signatures are due to an
additional quantum interference effect between the Autler-Townes
resonances. It is particularly important for the existence of EIT that the
dephasing rate of the quantum coherence~$\gamma_{\text{nr}}$ be much smaller than the dephasing rate of the polarization~$\gamma_{\text{probe}}$. An increased drive intensity
of $I_{\text{d}}=0.7\,\text{MW}/\text{cm}^{2}$ leads to a larger separation of the Autler-Townes resonances and a reduction of the peak slowdown factor $S'$, see Figs.~\ref{figure2}(c) and~(d). Also, an additional excitation-induced broadening of the spectral features for higher drive intensities occurs because the dephasing
contributions depend on the level occupations and the polarizations, so that
the dephasing of a particular transition depends on the drive pulse.

Keeping the drive intensity
at $I_{\text{d}}=0.25\,\text{MW}/\text{cm}^{2}$, but reducing the temperature to 150\,K leads to weaker dephasing and thus a more pronounced effect of the quantum interference, i.e., a more pronounced dip in Fig.~\ref{figure2}~(e) and higher peak slowdown in Fig.~\ref{figure2}~(f) compared with Figs.~\ref{figure2}~(a) and~(b), respectively. Due to the different occupation of the states
for lower temperatures the HF shift of the probe transition energy is
around 3.9\,meV for 150\,K, instead of 4.2\,meV for 300\,K. Also
the asymmetry of the spectra induced by HF corrections is more
pronounced for lower temperatures. 

\begin{figure}[tbp]
\centering
\includegraphics[trim=1.5cm 3.5cm 1.5cm 4.5cm,clip,scale=0.35,angle=0]{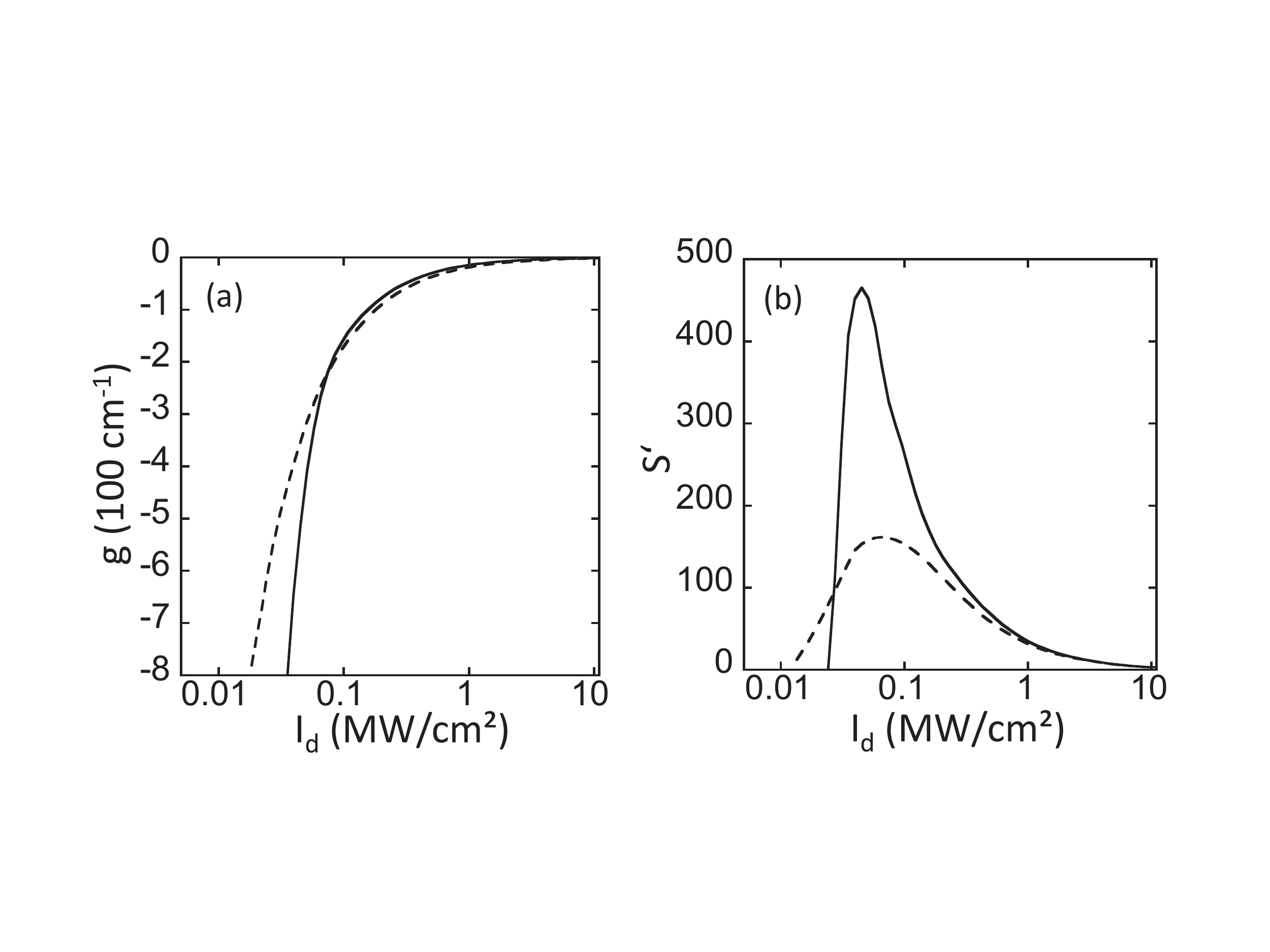}
\caption{Peak gain (a) and peak slowdown (b) versus drive intensity
  for a lattice temperature of 300~K (dashed line) and 150~K (solid
  line).}
\label{figure3}
\end{figure}

The dependence on the drive intensity for the gain and the peak slowdown is shown in Fig.~\ref{figure3} for lattice temperatures of 150\,K and 300\,K. For small drive intensities the peak gain and the peak slowdown increase with intensity because the effectiveness of the interference between the dressed states increases, which reduces the peak absorption. For drive intensities of about $I_{\text{d}}=0.1\,\text{MW}/\text{cm}^{2}$ and above the Autler-Townes splitting increases and thus reduces the peak absorption. In this case, the interference effects between the dressed states become less pronounced and peak slowdown decreases with increasing intensity. The peak gain still increases even when the interference becomes less effective because the Autler-Townes splitting continues to increase with intensity.

\begin{figure}[tbp]
\centering
\includegraphics[trim=4cm   4cm   4cm   4cm  ,clip,scale=0.5 ,angle=0]{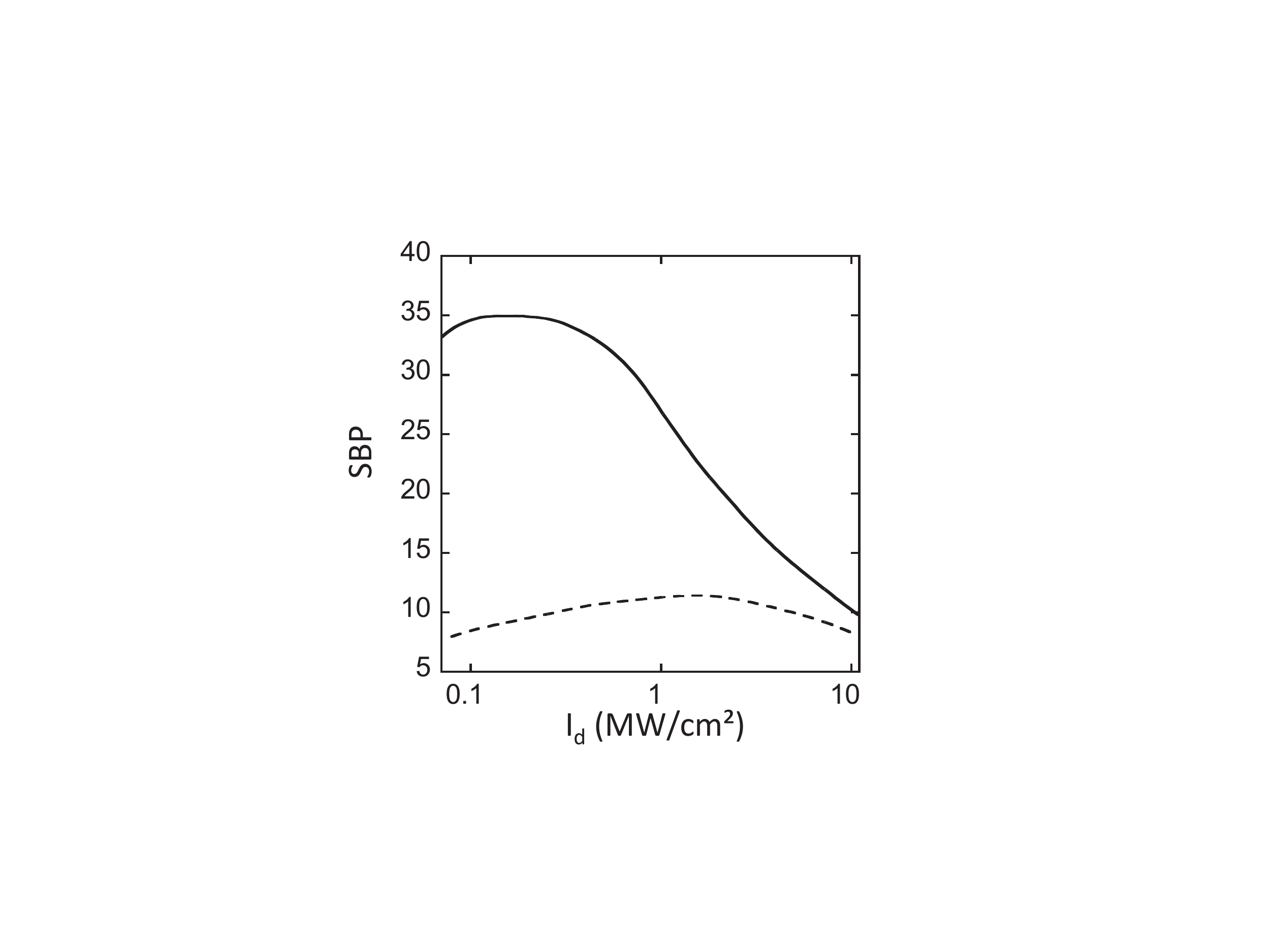}
\caption{Slowdown bandwidth product versus drive intensity for 
a lattice temperature of 300~K (dashed line) and 150~K (solid line).}
\label{figure4}
\end{figure}

The slowdown-bandwidth product (SBP) \cite{deng1} is an important
characteristic for the usefulness of quantum coherence schemes to slow
down pulses as already discussed in Ref.~\onlinecite{apl:quantum-coherence}. From
spectra such as Fig.~\ref{figure2} we obtain the SBP $\omega
_{1/2}n_{b}d\chi _{r}/d\omega _{p}$, where $\omega _{1/2}$ is the FWHM
of the h$_{0}^{\text{b} }\leftrightarrow \text{e}_{1}$
resonance. We investigate here the dependence of the SBP on the drive
intensity and the influence of the temperature as shown in figure
\ref{figure4} for $300$~K and $150$~K. Because the measurement of
the bandwidth for slow-down is only useful, if the Autler-Townes
splitting of the resonance is clearly visible, the product is not
calculated for low drive intensities. The increase of the
Autler-Townes splitting with increasing drive intensities leads to an increase of the SBP in a drive intensity range in which the slowdown $S'$ already decreases. For still higher intensities the pronounced drop in $S'$ wins over the increasing broadening. Further the
smaller broadening of spectral features for lower temperatures
influences the result by increasing the slow-down bandwidth. If the
slowdown-bandwidth product is compared to the one calculated in
Ref.~\onlinecite{apl:quantum-coherence} for a $\Lambda $-scheme, we have a
tremendous improvement. However, the improvement of the
slowdown-bandwidth product is even more pronounced for lower
temperatures. Encouraged by this promising result we investigated the
propagation of the probe pulse and calculated the slow-down factor for
various propagation conditions as presented in the following section.

\subsection{Slowdown factor due to pulse propagation\label{subsec:pulse-propagation}}

The analysis of the performance of the QD molecule for slowing
down light is extended by including the effects of propagation. Because in experiments and applications the
important information is in the spatio-temporal dynamics of the probe
pulse. 
The drive field is taken as cw or, for numerical reasons, as a pulse much longer than the probe. As shown in
Fig.~\ref{figure1a} and explained in the following its reasonable to neglect the
propagation of the drive field in growth direction while accounting for propagation
of the probe pulse in the plane of the well: 
The quantum well considered here, which includes the active region, has a width of
30\,nm in growth direction. Even if we assume a structure composed of
several quantum wells to achieve an increased confinement factor for the probe pulse, the total width of the
active region stays far below 1\,$\mu $m. Therefore propagation
effects for the drive pulse can be neglected and we concentrate on the
propagation of the probe pulse. We assume a quantum well with an
extension of 250\,$\mu$m in $y$-direction, in which the active region is
contained. Further we assume a long drive pulse of duration 200\,ps
and a spot radius larger than 250\,$\mu $m centered on the quantum
well. A probe pulse is initialized to occur in the middle of the drive
pulse; this defines the propagation distance zero. The
finite drive pulse duration is only introduced for numerical
reasons. In an experiment it could be a cw drive.

We compare the propagation results for a cosh$^{-2}$ probe pulse with
a FWHM of $17.6$~ps and $35.3$~ps with the results for a cw probe
field without propagation effects as described and calculated in the
previous section. The calculation of the gain and the slowdown factor
due to pulse propagation of the probe pulses is discussed below. The
FWHM is given for the field amplitude and corresponds to a FWHM of
$12.1$~ps and $24.2$~ps for the field intensity, respectively. The
calculation is done for a lattice temperature of $300$~K. We start the
probe pulse at $0$ $\mu $m with the relative time $t'_{0}=t_{0}$ (see
time transformation for slowly-varying Maxwell equations) and
propagate the probe pulse in the relative time $t^{\prime }$. After a
propagation length $d_{P}$ we determine the distance in the relative
time $\Delta t'$ between the initial and the propagated probe pulse
peak maximum and calculate the slow down factor $S-n_{b}$ averaged
over the propagation distance. The difference between the initial and
the propagated peak maximum of the probe pulse can be used to
calculate the amplitude gain of the probe pulse averaged over the
propagation distance.

\begin{figure}[tb]
\centering
\includegraphics[trim=1.5cm 3.5cm 1.5cm 4.5cm,clip,scale=0.35,angle=0]{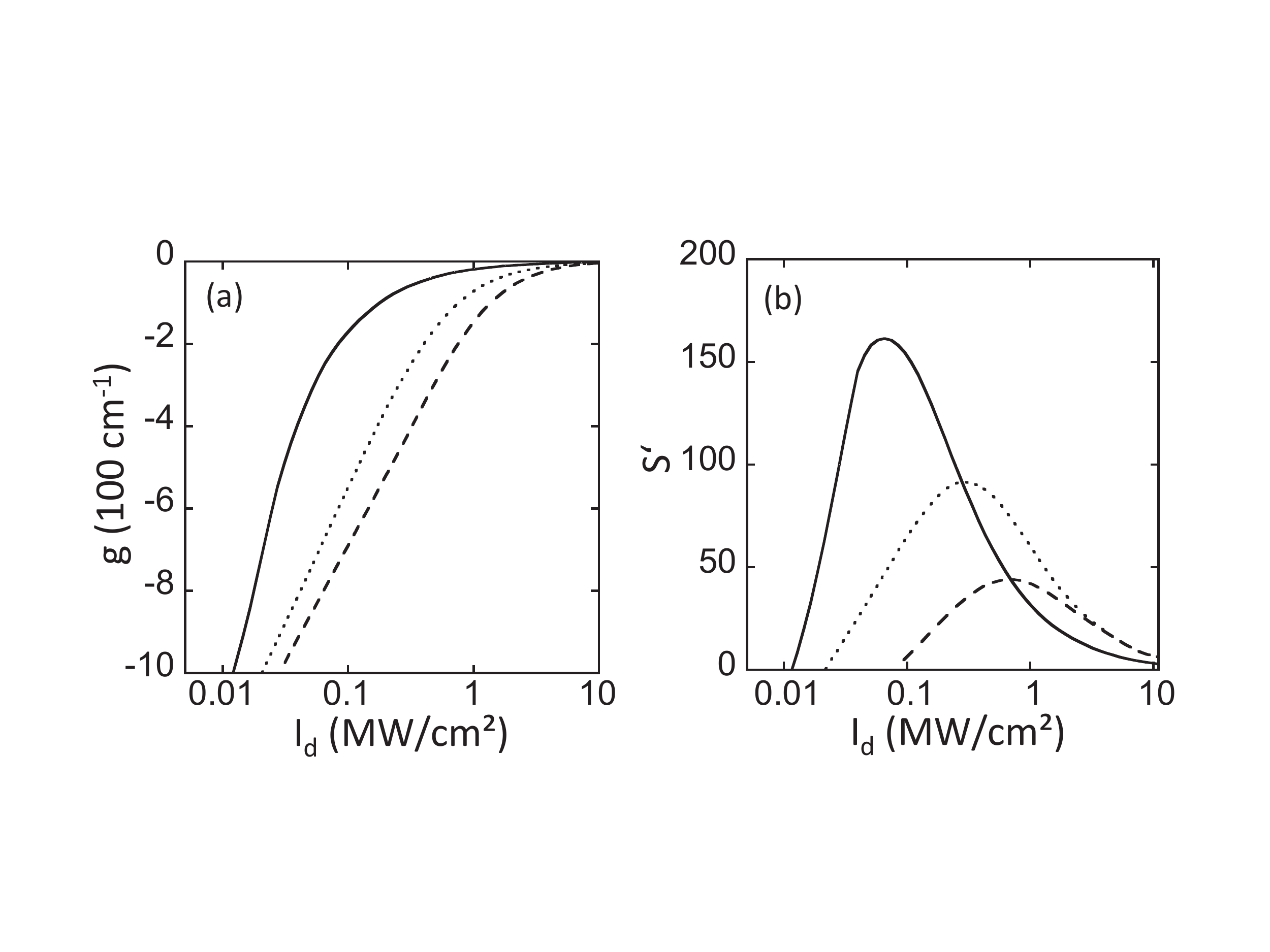}
\caption{Gain (a) and slowdown (b) versus drive pulse intensity after a
propagation distance of 1~$\protect\mu$m for a cosh$^{-2}$ probe pulse with a FWHM of 17.6~ps (dashed line) and
35.3~ps (dotted line) and a lattice temperature of 300~K. For comparison the
gain and slowdown factor for a cw probe (black solid line) without
propagation effects is plotted.}
\label{figure5}
\end{figure}

Figure \ref{figure5} shows the gain and slow-down factor
calculated for different drive pulse intensities and for a propagation
distance of $d_{P}=1$ $\mu $m. For the short probe
pulse compared to the long probe pulse and the cw probe field, the drive
pulse intensity has to be higher to reach a comparable transparency, so that
the dependence of the slow-down factor on the intensity is shifted and
damped. The reason for this behavior is that the polarization of the probe
pulse needs some time to build up the coherences that lead to the steady
state Autler-Townes splitting and, in turn, to EIT with
slow-down. Additionally, the gain and slow-down increase for longer
propagation distances. This can be explained with the increasing temporal
broadening of the probe pulse for longer propagation distances due to small
absorption effects: For the probe pulse with an initial FWHM of $17.6$~ps
propagation effects lead to a slight temporal broadening and a slightly
smaller (temporal) gradient of the pulse resulting in higher gain and
slowdown for the spatial propagation. For a probe pulse with an initial FWHM
of $35.3$~ps these effects are less pronounced due to the longer pulse with
smaller (temporal) gradient of the field.

\begin{figure}[tb]
\centering
\includegraphics[trim=1.5cm 3.5cm 1.5cm 4.5cm,clip,scale=0.35,angle=0]{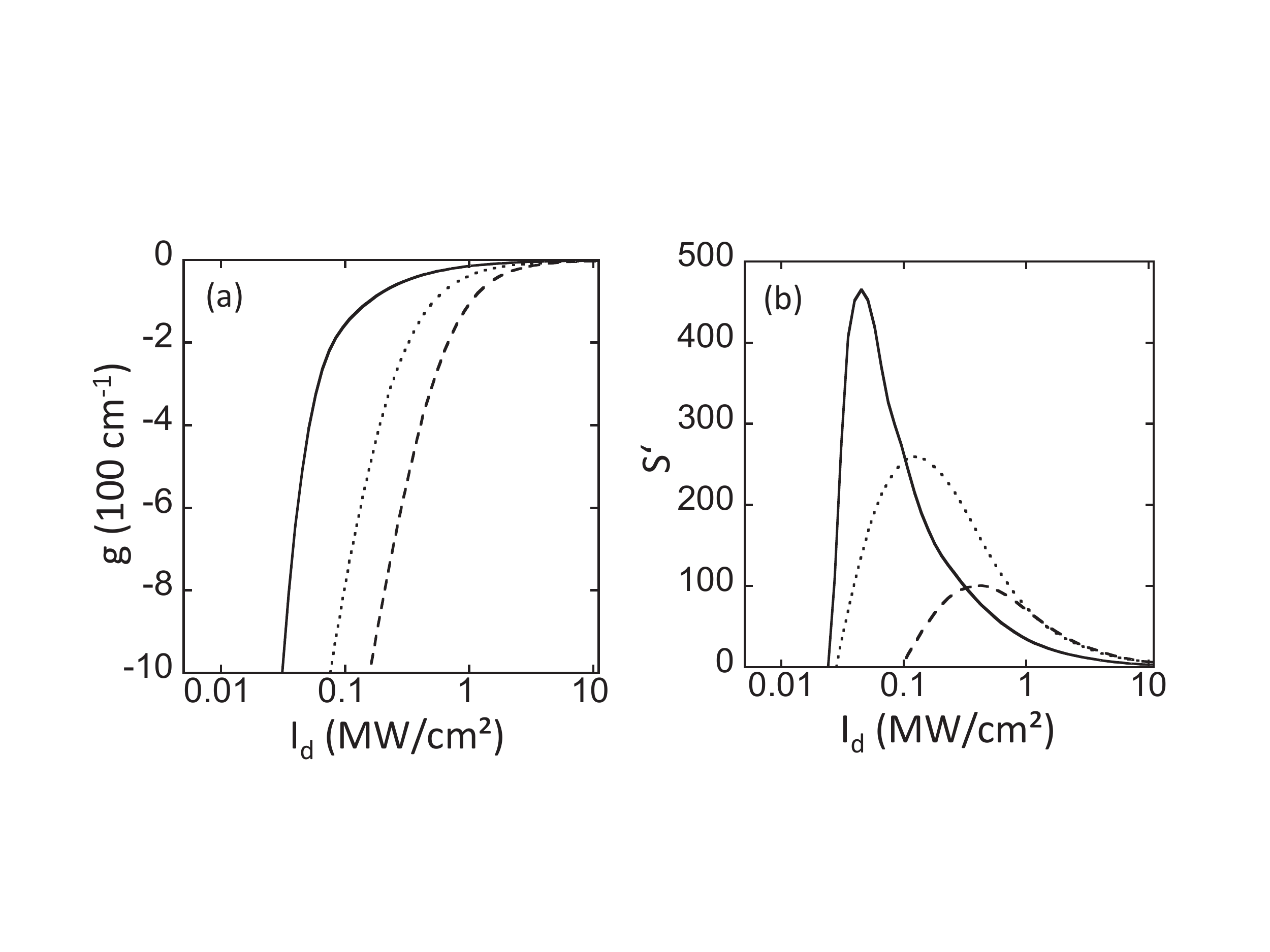}
\caption{Same plot as in Fig.~\protect\ref{figure5} for a lattice temperature of
  150~K.}
\label{figure6}
\end{figure}

\begin{figure}[tb]
\centering
\includegraphics[trim=2cm 1cm 3cm 3cm,clip,scale=0.4,angle=0]{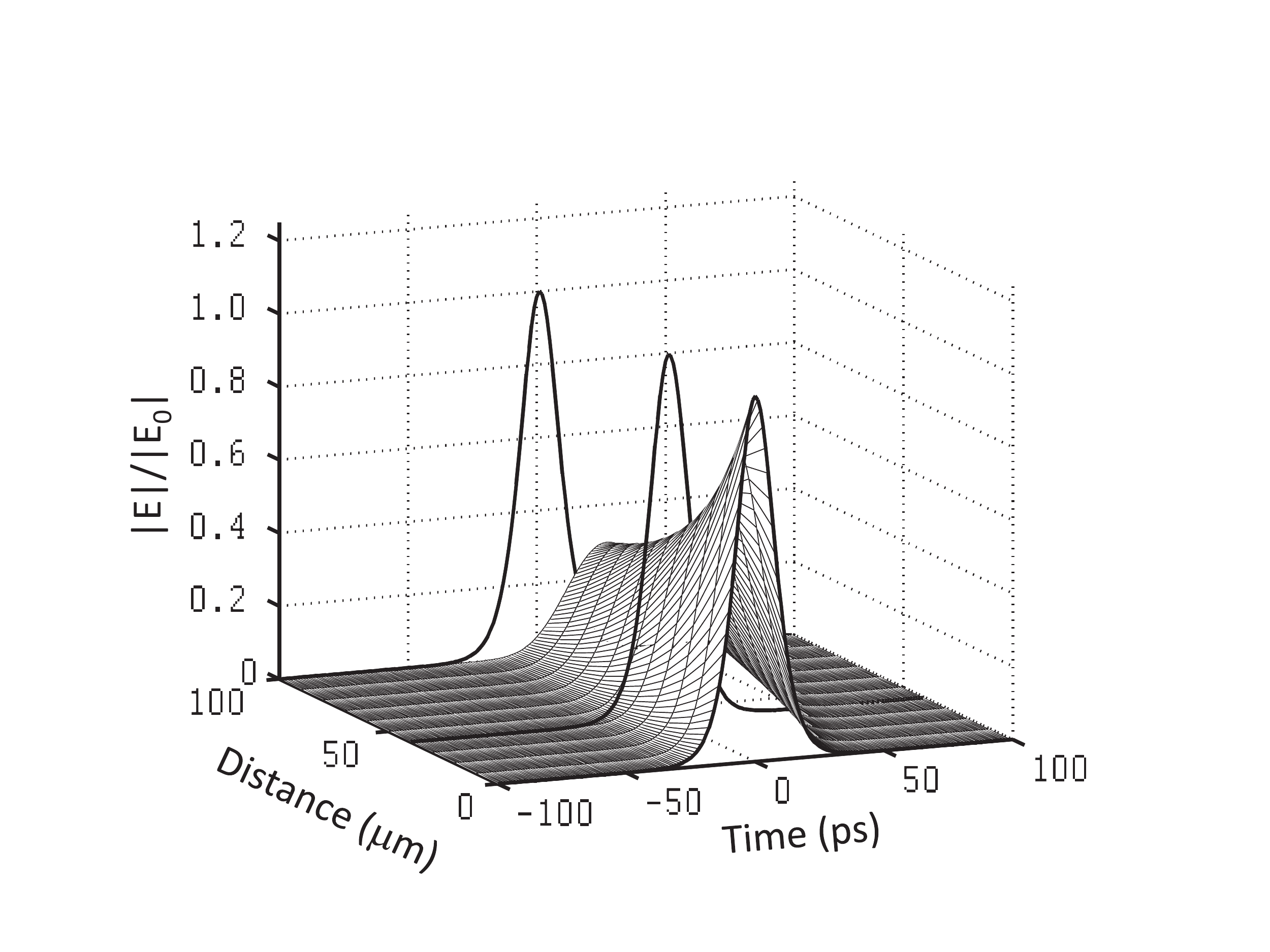}
\caption{Temporal shape of the probe pulse field versus propagation distance
  for a drive pulse intensity of $0.7$ MW cm$^{-2}$. The shape of the
  reference pulse versus propagation distance is plotted at $z=0\,\mu$m,
  $z=50\,\mu$m and $z=100\,\mu$m for comparison. The initial shape of the
  probe and reference pulse is a cosh$^{-2}$. The lattice temperature is 300~K.}
\label{figure7a}
\end{figure}

\begin{figure}[tb]
\centering
\includegraphics[trim=2cm 1cm 3cm 3cm,clip,scale=0.4,angle=0]{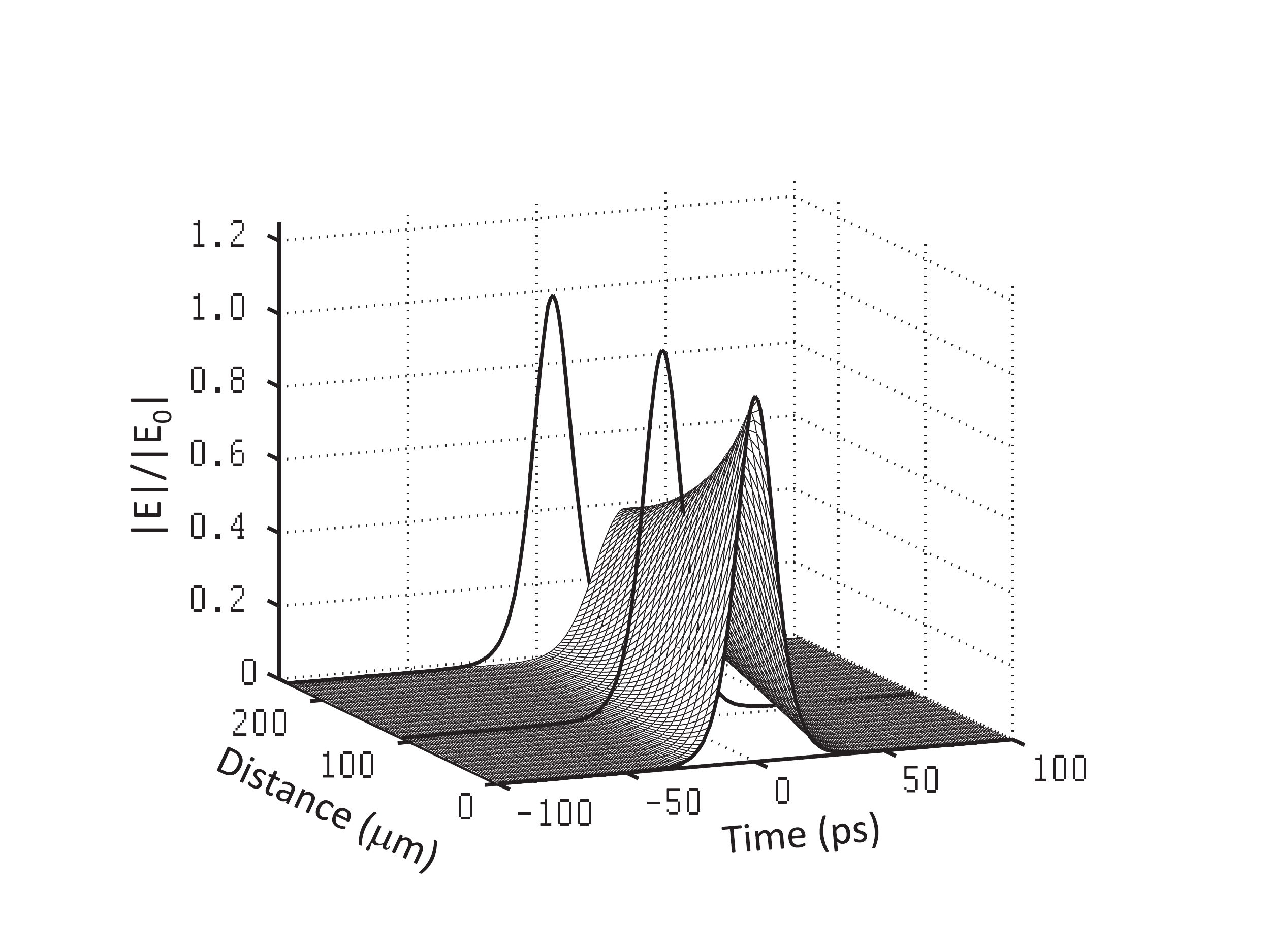}
\caption{Same as Fig.~\ref{figure7a} for a drive pulse intensity of $1.8$ MW
  cm$^{-2}$. The reference pulse is shown for $z=0\,\mu$m, $z=100\,\mu$m and $z=200\mu$m.}
\label{figure7b}
\end{figure}

\begin{figure}[tb]
\centering
\includegraphics[trim=2cm 1cm 3cm 3cm,clip,scale=0.4,angle=0]{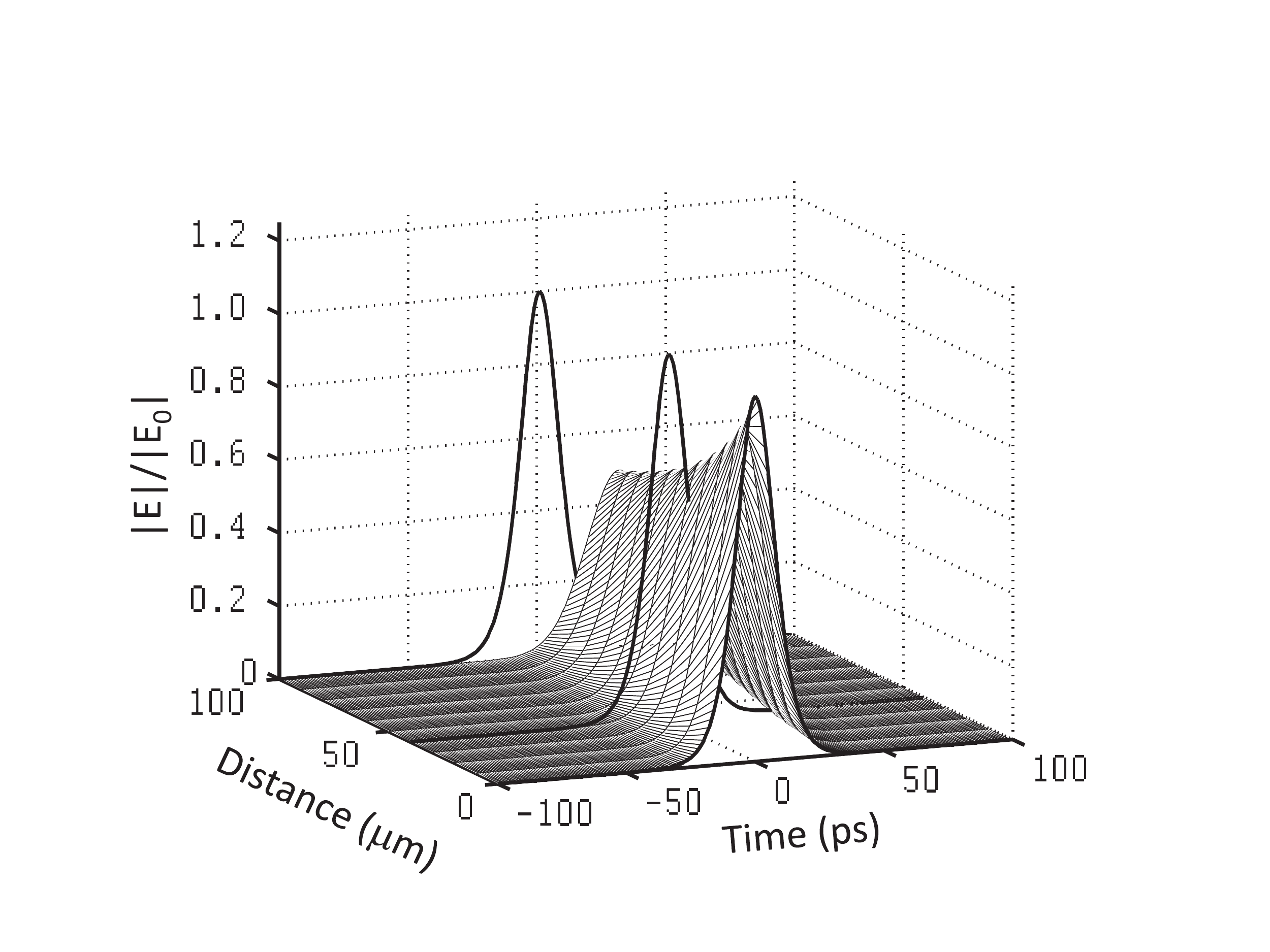}
\caption{Temporal shape of the probe pulse field versus propagation distance
  for a drive pulse intensity of $0.7$ MW cm$^{-2}$. The shape of the
  reference pulse versus propagation distance is plotted at $z=0\,\mu$m,
  $z=50\,\mu$m and $z=100\,\mu$m for comparison. The initial shape of the
  probe and reference pulse is a cosh$^{-2}$. The lattice temperature is 150\,K.}
\label{figure8a}
\end{figure}

\begin{figure}[tb]
\centering
\includegraphics[trim=2cm 1cm 3cm 3cm,clip,scale=0.4,angle=0]{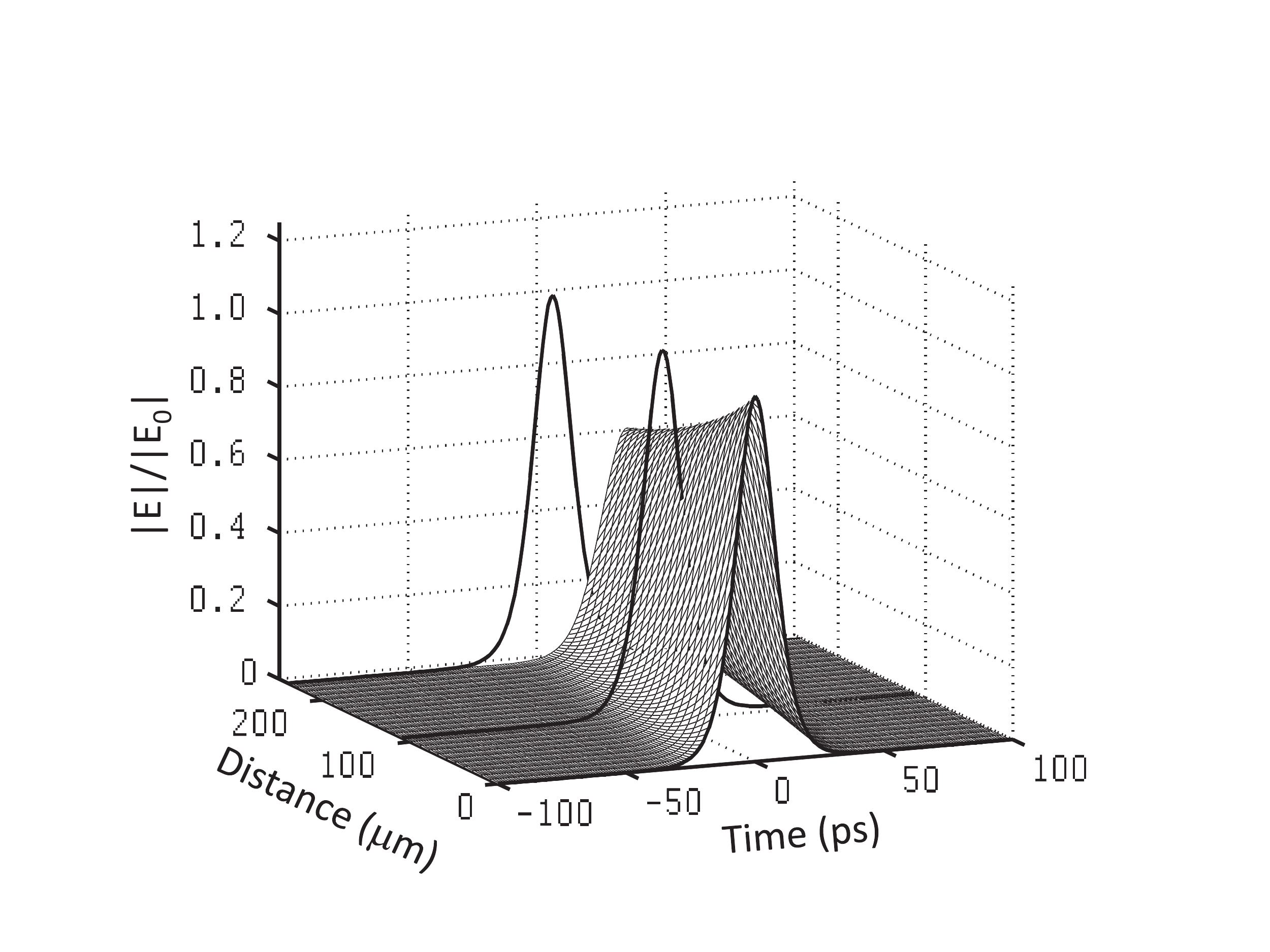}
\caption{Same as Fig.~\ref{figure8a} for a drive pulse intensity of $1.8$ MW
  cm$^{-2}$. The reference pulse is shown for $z=0\,\mu$m, $z=100\,\mu$m, and $z=200\,\mu$m.}
\label{figure8b}
\end{figure}

In Figs.~\ref{figure7a} - \ref{figure8b}, we compare 
the shape of the probe pulse after different propagation distances for
different drive pulse intensities and lattice temperatures.
The temporal shape of the probe pulse is shown in \emph{real time} for propagation distances of a
few $100\,\mu$m. For comparison a reference pulse, i.e., a propagated pulse shape without slowdown,
is also plotted. This reference pulse is obtained by propagating the probe
field in the host material with refractive index $n_{b}$, but without the QD
molecules. A temporal shift of the probe
pulse peak against the reference pulse peak to positive times after a spatial propagation corresponds
to a slow-down of the probe pulse and a temporal shift to negative times
corresponds to a speed-up of the probe pulse. Fig.~\ref{figure7a} contains
results for a drive pulse intensity of $0.7$~$\text{MW/cm}^{2}$ and a maximum
propagation distance of $100$ $\mu$m. The slow-down of the
probe pulse is clearly visible and the probe pulse peak after a propagation
distance of $100$ $\mu$m already has a noticeable separation to the initial probe
pulse peak with a moderate loss of amplitude and only a small distortion.
Fig.~\ref{figure7b} changes the drive pulse intensity to
$1.8$~$\text{MW/cm}^{2}$. In this case, a pulse separation is reached for longer propagation
distance, i.e., less efficient slowdown, but with a lower loss of amplitude and a smaller
distortion. Therefore, the shape of the probe pulse is plotted for
propagation distances up to $250$ $\mu$m. The lower loss of amplitude and the lower distortion is due to the lower
absorption at higher intensities already visible in Fig.~\ref{figure5}. 
Therefore, the absorption, i.e., distortion,
and the slowdown factor of the probe pulse have to be balanced to obtain decent
results for a slow light application. Here, for both drive pulse intensities a separation between the
initial peak and the propagated peak of the probe pulse is possible with a
moderate loss of amplitude.

\begin{figure}[tb]
\centering
\includegraphics[trim=5cm 5cm 7cm 3cm,clip,scale=0.5,angle=0]{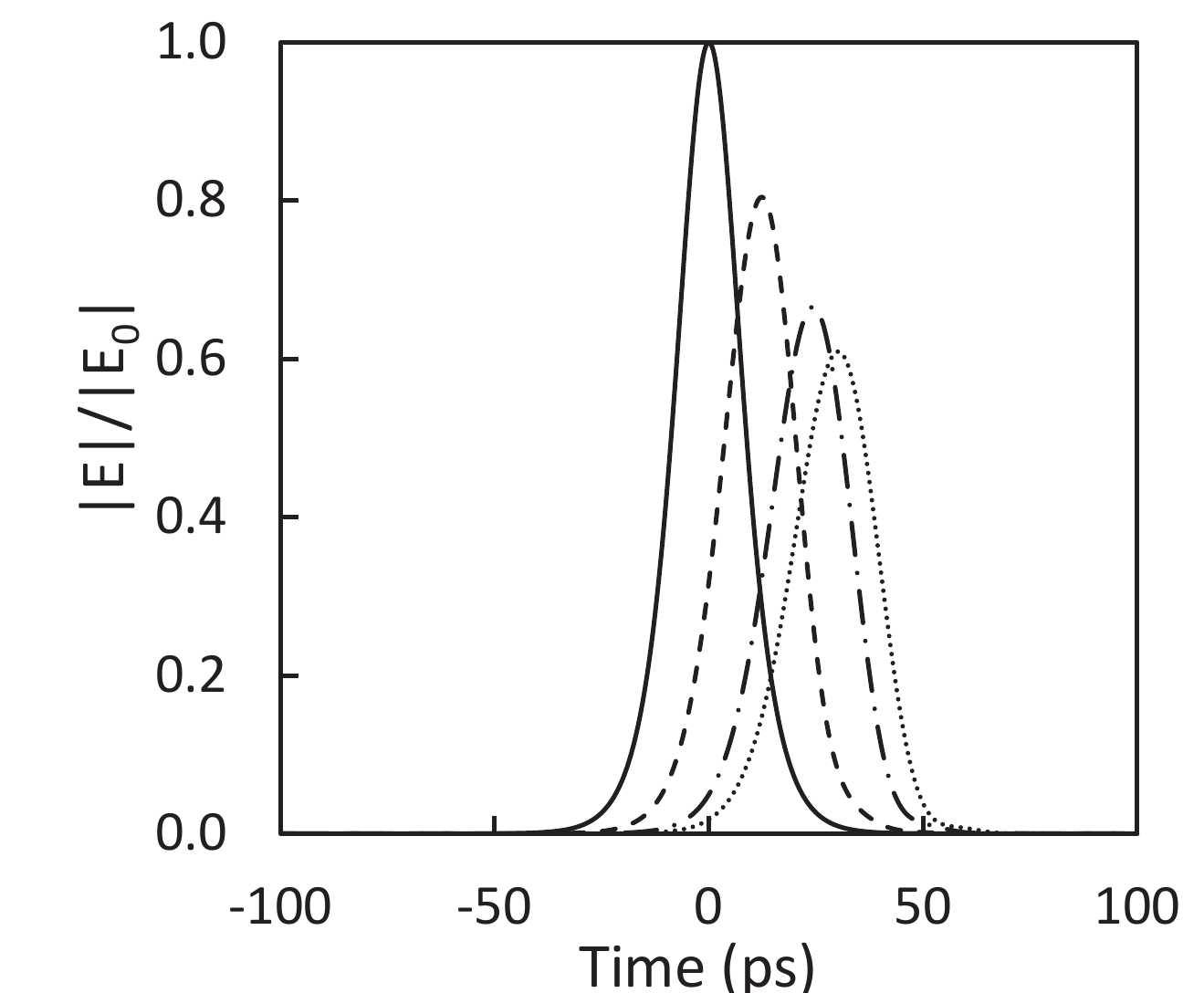}
\caption{Temporal field shape of the probe pulse for a drive pulse
  intensity of 1.8~MW/cm$^{2}$ after
  propagation distance of 0~$\mu$m (solid), 100~$\mu$m (dashed),
  200~$\mu$m (dashed-dotted), 250~$\mu$m (dotted). The initial shape of the
  probe pulse is a cosh$^{-2}$. The lattice temperature is 150~K.}
\label{figure9}
\end{figure}

We also calculated the propagation for a cosh$^{-2}$ probe
pulse with a FWHM of $17.6$~ps and a cosh$^{-2}$ probe pulse with a FWHM of $
35.3$~ps for a lattice temperature of $150$~K. The gain and slow-down factor
are calculated for a set of drive pulse intensities and for a propagation
distance of $d_{P}=1$ $\mu $m and shown in Fig.~\ref{figure6}. Furthermore the cw probe result is
plotted for comparison. We obtain for gain and slowdown factor plots
generally the same qualitative behavior as described for the case of $300$~K
(see figure \ref{figure5}), but better results, i.e., less absorption and
more pronounced slow-down. 
This improvement is also evident from the comparison of
the probe pulse shape between different propagation distances in
Figs. \ref{figure8a} and \ref{figure8b}: For a drive pulse intensity
of $0.7$~$\text{MW/cm}^{2}$ and $1.8$~$\text{MW/cm}^{2}$, the shape of the
probe pulses is plotted after propagation distances up to $100\,\mu$m and $250\,\mu $m, respectively. Again we obtain the same qualitative
behavior between the two drive pulse intensities as already analyzed for the
case of $300$~K. In addition, for a lattice temperature of $150$~K less
absorption and higher slow-down are obtained and thus a pulse separation
with a smaller loss of amplitude and a smaller pulse distortion can be
realized.
In Fig.~\ref{figure9} the results of Fig.~\ref{figure8b} are shown as a two
dimensional graph, for better quantitative comparison. Now, the shape of the
probe pulse is shown vs. relative time after a propagation
distance of $0\,\mu $m, $100\,\mu$m, $200\,\mu $m and $250\,\mu $m. Because the shape of the probe pulse is plotted
against the relative time, a temporal shift of the probe
pulse peak to positive relative times after a spatial propagation corresponds
to a slow-down of the probe pulse and a temporal shift to negative relative
times corresponds to a speed-up of the probe pulse. 
Fig.~\ref{figure9} shows that the separation of the pulse peaks is about 29\,ps
after $250\,\mu$m with a small loss of amplitude and negligible distortion. 
These numerical values are promising for slow light applications.  

To conclude this section, we would like to give some numbers regarding effective dephasing rates for our QD molecule system and setup.
We determined polarization dephasing rates for three transitions: the quantum coherence $e_0\leftrightarrow e_1$ , the probe transition $h_0\leftrightarrow e_1$ and the drive transition $h_0\leftrightarrow e_0$. As in the calculations for Figs.~\ref{figure7a} - \ref{figure8b}, we used an extremely long drive pulse, but also a long probe pulse to mimic cw excitation conditions. The pulse frequencies $\omega_d$ and $\omega_p$ were chosen to be resonant to the renormalized transition energies corresponding to these excitation conditions. We then fit polarization decay rates $\gamma_{e_0,e_1}$, $\gamma_{h_0,e_1}$, $\gamma_{h_0,e_0}$ to the polarization dephasing contributions $S_{e_0,e_1}$, $S_{h_0,e_1}$, and $S_{h_0,e_0}$, respectively. Typical values of the dephasing for the
quantum coherence are around $\gamma_{e_0,e_1}\equiv \gamma_{\text{nr}}\approx 0.01\,\text{ps}^{-1}$ 
and the dephasing on the drive and probe transitions is generally on the order 
of $\gamma\approx 0.5\,\text{ps}^{-1}$ for the range of drive intensities 
and temperatures (150\,K and 300\,K) considered here. The dephasing rates of the drive and probe transitions are typical of QDs,
whereas the comparatively small dephasing rate of the quantum coherence is due
to the design of the QD molecule and the chosen setup. The dephasing of the quantum coherence in the
QD molecule obtained from our microscopic calculation
at and above $150$~K is still quite a bit slower than the dephasing rate
$0.125\,\text{ns}^{-1}$ assumed in
a recent AMO-like calculation of EIT based slow-light in 
single QDs.~\cite{mork_JOSAB} Such a slow dephasing along with lifetime-limited 
linewidths, which were also assumed in Ref.~\onlinecite{mork_JOSAB}, are generally only realized at very low temperatures.


\section{Comparison to a single QD\label{sec:results_single}}

Here we wish to compare the results for pulse slowdown in the optimized QD
molecule with earlier results on QDs. First, in the $\Lambda$ schemes for an
ensemble of single QDs, as investigated in our earlier
Refs.~\onlinecite{qcpinsqd,apl:quantum-coherence,jmo}, the quantum coherence
connects two hole states and is therefore susceptible to the same
dephasing contributions as the drive or probe (electron-hole) polarization, where
the dominant contribution of the dephasing comes from the hole states because
they are closely spaced and because of the polaronic broadening the
electron-phonon interaction can efficiently couple them. In this case, a short
drive pulse is necessary to slow down the probe
pulse,~\cite{apl:quantum-coherence,jmo} but the time window during which the
probe pulse is slowed down, is too short.~\cite{dissertation} To highlight the
slowdown achievable in QD molecules with cw drive fields we would like to
compare them with single QDs for the same quantum coherence scheme, namely a
$V$ scheme. Our choice of
$V$ scheme is also supported by investigations of quantum coherence schemes in
single QDs which found that the structural QD parameters can generally be more
easily optimized for $V$ schemes than for other
schemes.~\cite{Nielsen1,Nielsen2} In the following we
investigate the pulse slowdown for $V$ schemes using single QDs in the same
manner as for the QD molecules. 
For the purpose of this section, it is not necessary to design novel QD
molecule structures using finite model potentials and wavefunctions that are
checked against k$\cdot $p-calculations. Instead, for simplicity, we work with
a simpler QD model with a harmonic oscillator confinement potential. This
model was used, e.g., in Ref.~\onlinecite{qcpinsqd}.

\subsection{Single QD model for a $V$ scheme\label{subsec:results_single_model}}

The QD model and the calculation of the pulse slowdown for $V$
schemes is similar between single QDs and QD molecules.
Here, we only highlight the differences.
We assume cylindrical single QDs described by the Hamiltonian
(\ref{3DSchroedinger_cylinderQD}) in
envelope approximation, but without the finite potential (\ref{3DPotential_cylinderQD}).
Instead, we replace the in-plane confinement potential in equation
(\ref{2DSchroedinger_cylinderQD}) with a harmonic oscillator confinement potential.
This is a good approximation, because
measurements of the dependence of the lowest bound states in a QD are
also in agreement with a spectrum of a harmonical oscillator.~\cite{qcpinsqd} The
in-plane Hamiltonian can be solved by separating the radial and angular
dependence using Hermite polynomials as also described in Ref.~\onlinecite{qcpinsqd}. This
approximation would be inappropriate for QD molecules, because the
determination of wave functions and energy levels from a \emph{finite} confinement potential for each single QD
is necessary to calculate the wave functions and energy levels of the
electronically coupled QDs (i.e., QD molecules). 

\begin{figure}[tb]
\centering
\includegraphics[trim=6cm 2cm 4.5cm 4cm,clip,scale=0.5,angle=0]{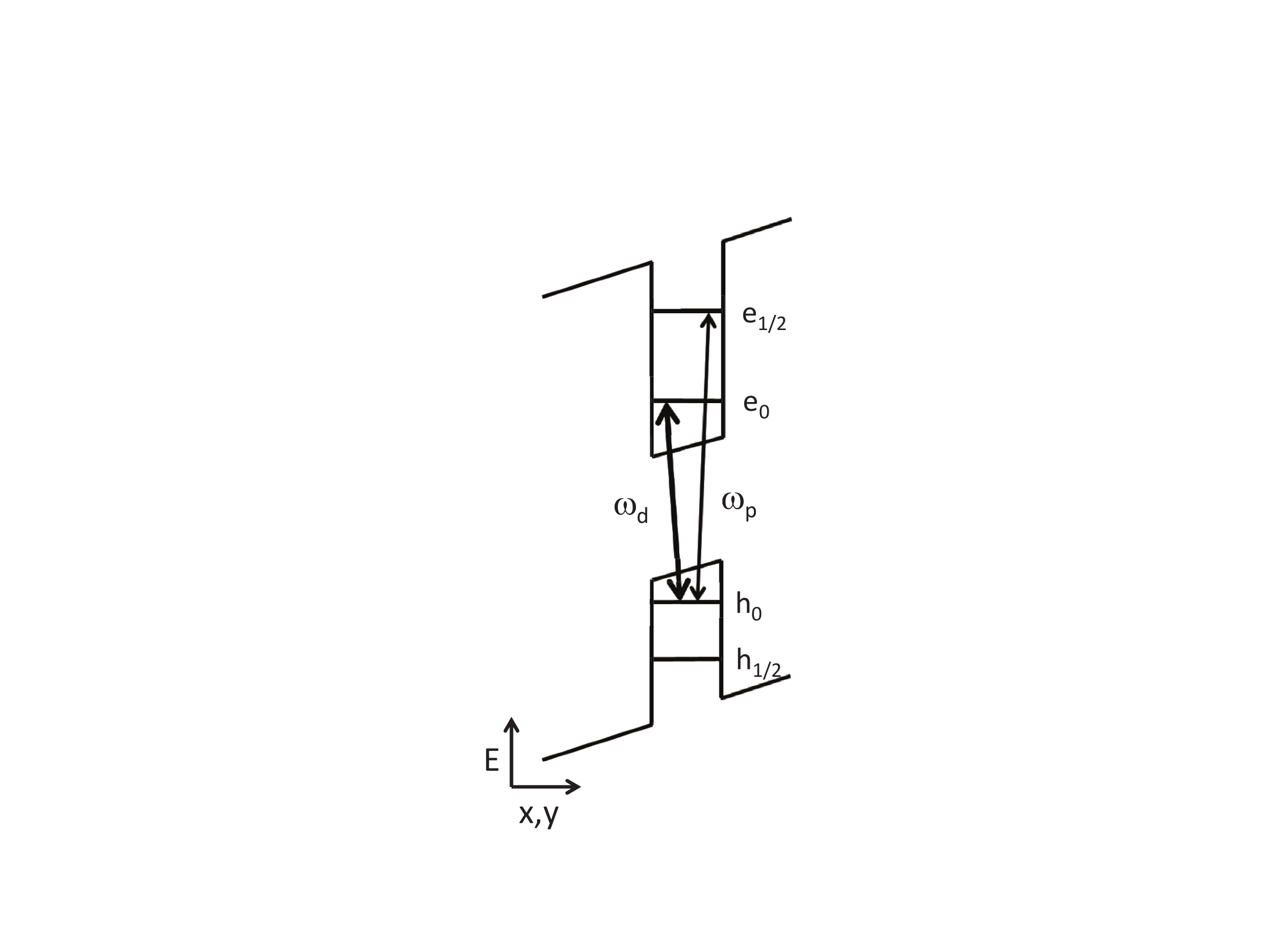} 
\caption{Schematic picture of the geometry
  of the single QD described in the text.
The resonant probe and drive fields in a $V$-type quantum coherence scheme are also shown.}
\label{figure10_1}
\end{figure}

We assume an ensemble of InGaAs-based QDs embedded in a GaAs quantum well with a
width of 16\,nm, which leads to three
confined electron and hole states. Thus we have one doubly degenerate excited
state and one ground state with the energy values in table~\ref{table-2}. The
line-up of the levels is shown schematically in Fig. \ref{figure10_1}. 
Using an analytical model of cylindrical QDs only diagonal transitions are
dipole allowed because of symmetry considerations. However, to realize a $V$
scheme one needs off-diagonal interband transitions. We achieve this by
including a symmetry breaking static electric field. To make off-diagonal
dipole matrix elements appreciable, we use an external electric field in the
plane of the quantum well with a field strength of $4.0$ mV nm$^{-1}$. The
calculated dipole matrix elements make a $V$ scheme with a drive pulse between
the electron and hole ground state and a probe pulse between the hole ground
state and the excited electron states possible. The quantum coherence of
the $V$ scheme is between the electron ground and the excited electron
states. The energy gap between the electron and hole ground state is taken
to be $1.2$\,eV. 

As shown in table~\ref{table-2}, we assume a \emph{deep} confinement of the QD
states. This choice is necessary to obtain noticeable slowdown in our single QD $V$ scheme setup: 
As already explained in section \ref{sec:corcontr} the dominant dephasing
processes of our single QD $V$ scheme setup are those carrier-phonon dephasing processes which are associated with
real carrier-phonon scattering transitions. For a shallow confinement of the QD states the
hole-intersubband and the electron-intersubband contributions would be of equal
size due to the small energy spacing of the hole and the electron states.\cite{dissertation} For
a deep confinement of the QD states the energy spacing of the electron states is high enough to
suppress the electron-intersubband contribution. Thus, the carrier-phonon
dephasing of the quantum coherence is small compared to the
carrier-phonon dephasing of the probe polarization for a deep QD, but for a
shallow QD the two carrier-phonon dephasing rates would be of equal size. 

\begin{table}[bt]
\begin{tabular}{||l|c||l|c||}
\hline\hline
state & $E_e$ (meV)&state& $E_h$(meV) \\ 
\hline
e$_{0}$ & $-150$ & h$_{0}$ & $50$  \\
e$_{1/2}$ & $-60$ & h$_{1/2}$ & $20$\\ 
\hline\hline
\end{tabular}
\caption{Electron (e) and hole (h) energies of single-particle states
  in the single QD.}
\label{table-2}
\end{table}

\subsection{Numerical results for the single QD $V$ scheme\label{subsec:results_V_single}}

\begin{figure}[tb]
\centering
\includegraphics[trim=1.5cm 3.5cm 1.5cm 4.5cm,clip,scale=0.35,angle=0]{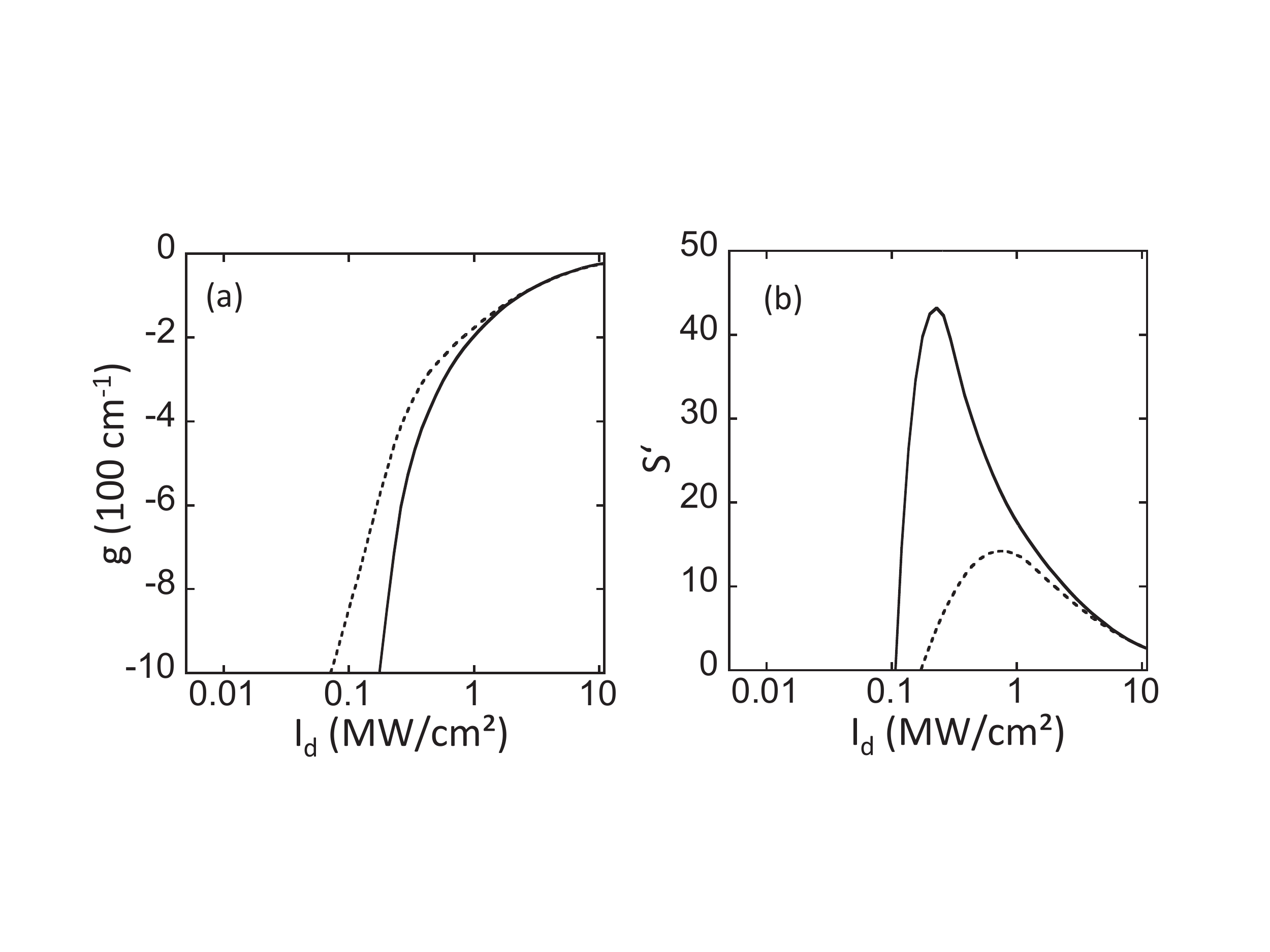}
\caption{Peak gain (a) and peak slowdown (b) versus drive intensity for a
  lattice temperature of 150~K (solid line) and 300~K (dashed line) for a
  deep single QD.}
\label{figure10_2}
\end{figure}

For the single QD we first investigate peak gain and peak slowdown as done in
figure~\ref{figure3} by comparing a lattice temperature of $150$~K with a lattice
temperature of 300\,K. 
For the calculation of the gain and the slow down factor, we use
the semiconductor Bloch equations (\ref{p-ab})--(\ref{n-b}) including a microscopic scattering and dephasing
contribution as described in Sec.~\ref{sec:corcontr}. The peak gain and peak
slowdown vs.\ drive intensity are shown in figure \ref{figure10_2}.
Below a drive intensity of 0.1\,MW/cm$^{2}$ we
find a significant peak absorption without peak-slowdown for both lattice temperatures.
Above a drive intensity of 0.1\,MW/cm$^{2}$ the peak
slowdown factor for similar peak absorption values is higher for lower
temperatures. This result again is obtained because the average phonon occupation is reduced for
lower temperatures, and a smaller carrier-phonon dephasing rate results for all
polarizations. This reduction is proportionally less pronounced for the
\emph{interband} and proportionally more pronounced for the \emph{quantum coherence}. The
dephasing processes of the quantum coherence which are associated
with real carrier-phonon scattering transitions are significantly
reduced. These processes no longer dominate over carrier-carrier and
pure-dephasing carrier-phonon processes of the quantum coherence. 
But an effectively long dephasing time for (realistic) slow light applications is still
not reached.

\subsection{Comparison between a $V$ scheme in a single QD and a QD molecule\label{subsec:results_single_comparison}}

\begin{figure}[tb]
\centering
\includegraphics[trim=1.5cm 3.5cm 1.5cm 4.5cm,clip,scale=0.35,angle=0]{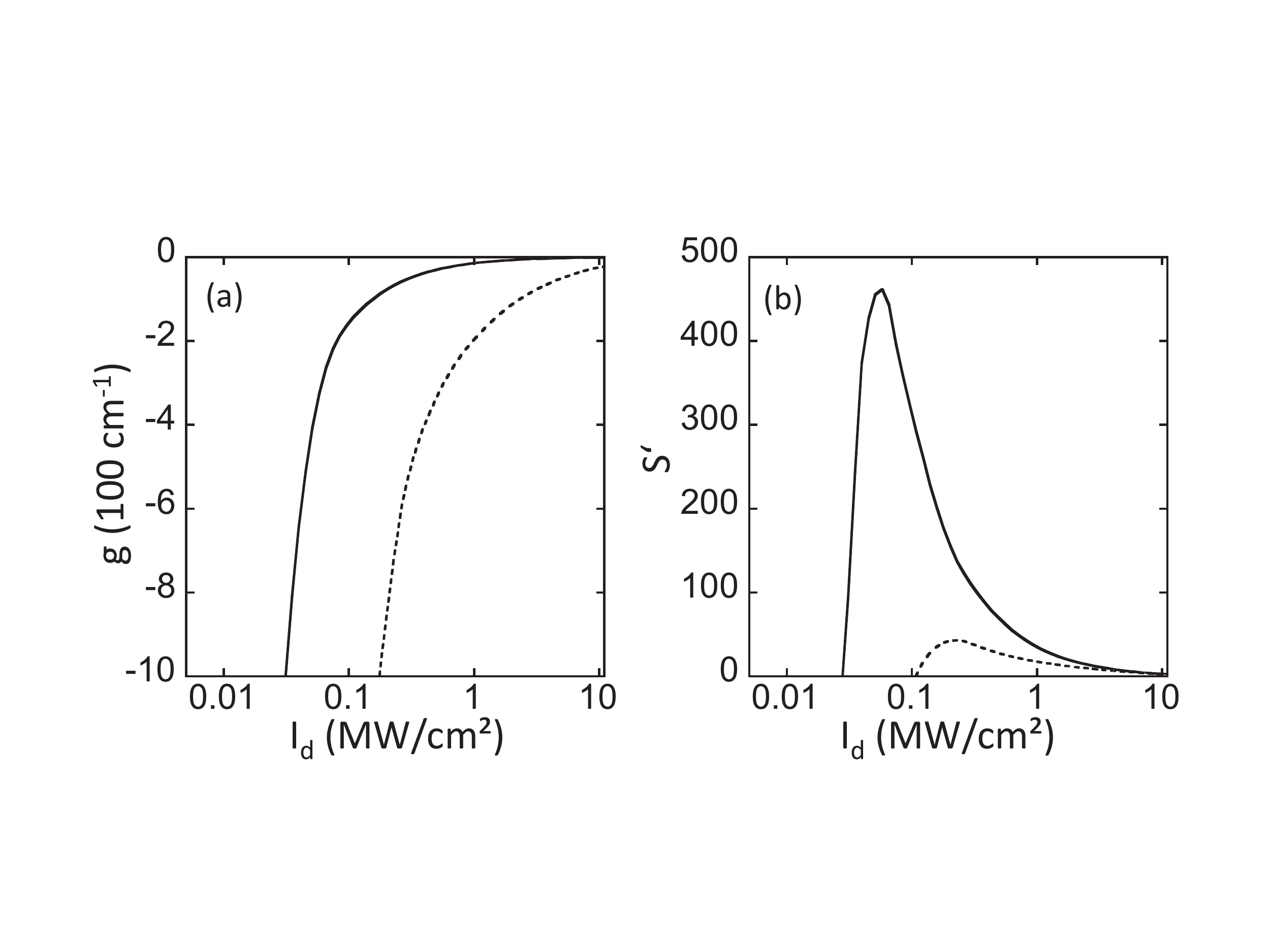}
\caption{Peak gain (a) and peak slowdown (b) versus drive intensity for the
  QD molecule (solid line) and the deep single QD. The
  lattice temperature is 150~K.}
\label{figure10_3}
\end{figure}

Finally we compare the results of the QD molecule and the deep single
QD for a lattice temperature of $150$~K. In figure \ref{figure10_3} the peak gain
and peak slowdown versus drive intensity is plotted for both setups. A
tremendous improvement of the peak slowdown factor for similar peak
absorptions values for the QD molecule compared to the deep single
QD is visible. This improvement can be explained in the following
way: The negligible wave-function overlap between the states of the
e$_{0}\leftrightarrow\text{e}_{1}$ transition of the $V$ system in the QD molecule has a
huge influence on the electron-phonon \emph{and electron-electron} dephasing
contributions of the quantum coherence. This influence reduces the
dephasing rate of the quantum coherence much more than the
dephasing rate of the interband probe polarization. Therefore, the comparison between the results of single QDs and
QD molecules shows, that an effectively long dephasing time can
only be engineered by using suitable QD \emph{molecules}. 

\section{Conclusion\label{sec:conclusion}}

In this paper we presented a microscopic analysis of quantum coherence
schemes, in particular electromagnetically induced transparency and group-velocity slowdown, in a special double-QD molecule design. We incorporated
scattering and dephasing effects, including polaronic effects in QDs, into the equations
of motion for the relevant polarizations and distribution functions. We used a
quasi-analytic model for QD single-particle states with parameters adjusted to
the results of $k\cdot p$-calculations for a realistic InGaAs-based QD. The
design of the double-QD molecule was geared towards achieving a long-lived
quantum coherence in a $V$ scheme involving two electronic levels localized at the individual QDs and a delocalized hole level. Starting from the quasi-analytic
QD model we constructed the states of the QD molecule and used these as input
in the equations of motion for polarizations and distribution
functions. Choosing probe and drive fields suitable for a $V$ scheme consisting of the
delocalized hole level and the two localized electron levels, we found cw slowdown
factors and slowdown-bandwidth products of the QD molecule that are far better
than our previous results on~$\Lambda$ schemes or results achieved by $V$
schemes in single QDs as presented in Sec.~\ref{sec:results_single}. We further combined the
microscopic material equations with a numerical calculation of the propagating
probe pulse and showed that a clear separation of the slowed down pulse with a
reference pulse can be achieved over distances of a few $\mu m$ with
acceptable pulse distortion and absorption. We emphasize that this result was made possible by 
the design of the QD molecule that yields a comparatively long dephasing time
on the quantum coherence. Importantly, the dephasing contributions that largely determine the figures of merit for the slowdown were not taken as constant rates, but arose from the microscopic treatment of the underlying interaction processes for a sufficiently realistic QD model.

\begin{acknowledgments}
This work was supported in part by Sandia's Solid-State Lighting Science
Center, an Energy Frontier Research Center (EFRC) funded by the US Department
of Energy, Office of Science, Office of Basic Energy Sciences. 
\end{acknowledgments}

\appendix 

\section{Electronic structure of QD molecules\label{QD-estructure}}

This appendix gives details of our calculation of the wave functions and
energies of QD molecules.  Since we are interested in the qualitative
properties of QD molecules, and wish to be able to easily describe the
structural parameters (i.e. different QD sizes and distances between the two QDs),
we use a simple and semi-analytical approximation for the description of the
QDs contained in the QD molecules. We stress that no band-mixing due to strain or piezoelectric effects are included, but the parameters used in the model have been chosen to compare well to a QD calculated by $k\cdot p $-theory. 


\subsection{Electronic structure of a cylindrical QD \label{app_dot}}

We assume a cylindrical QD with a confinement
potential of depth. For the Hamiltonian of the cylindrical QD in envelope
approximation we use%
\begin{equation}
H=-\frac{\hbar ^{2}}{2m}\nabla^2 +V(r,z)
\label{3DSchroedinger_cylinderQD}
\end{equation}
where the Laplacian~$\nabla^2$ and
\begin{equation}
V(r,z)=\left\{ 
\begin{array}{c}
0 \\  -V_{0}
\end{array}
\right. \left. 
\begin{array}{c}
\text{for }\left\vert z\right\vert >a \\ 
\text{for }\left\vert z\right\vert <a%
\end{array}%
\right. \left. 
\begin{array}{c}
\text{or} \\ 
\text{and}%
\end{array}%
\right. \left. 
\begin{array}{c}
\left\vert r\right\vert >b \\ 
\left\vert r\right\vert <b%
\end{array}%
\right.
\label{3DPotential_cylinderQD}
\end{equation}%
are expressed in cylindrical coordinates. The heights of the QD in $z$ 
direction is $h=2a$, the diameter is $2b$, and $V_{0}$ is the depth of
the confinement potential. The wave function $\Phi _{3D}$ of the
Schroedinger equation
\begin{equation}
H\Phi _{3D}=E\Phi _{3D}
\end{equation}%
has to be understood as an envelope function and $m$ as an effective mass.

We assume that the height is much smaller than the diameter of the QD.
Therefore the electrons and holes are strongly localized in the growth
direction~$z$. If we assume the separability of the wave function for the
in-plane and the $z$~direction, the three dimensional Schroedinger equation
reduces to a two and an one dimensional problem. 
Thus we can write for the
wave function
\begin{equation}
\Phi_{3D}(r,\varphi ,z)=N\, \Phi _{\parallel }(r,\varphi ) \Phi_{\perp }(z)
\end{equation}
where $N$ is a normalization constant.
Furthermore we used in $z$ direction
$\int \Phi_{\perp }^{*}(z) V(r,z) \Phi_{\perp }(z) \text{ d}z \approx -V_{0} \Theta(b-|r|) $
and the corresponding approximation for the in-plane direction 
to obtain a self-consistent set of equations.

For the Schroedinger equation in $z$ direction we have
\begin{equation}
\left[ -\frac{\hbar ^{2}}{2m}\frac{\partial^{2}}{\partial z^{2}}
+\tilde{V}_{\perp }\,\Theta(a-|z|)\right] \Phi_{\perp }(z)=E_{\perp }\Phi _{\perp }(z)
\end{equation}
Here, $\tilde{V}_{\perp}=-V_0 + T_{\|}$ is an effective one-dimensional potential that contains
the in-plane kinetic energy $T_\|=\tilde{V}_\|-E_\|$ which, in turn, depends
on the in-plane eigen-energy~$E_\|$. 
Since these kinetic energies $T_{\|}$ are
not known, we use an iteration procedure to calculate the in-plane and $z$
eigen-energies. We start the iteration by setting $\tilde{V}_{\perp}$
equal to $-V_0$. For the solution
we obtain for the symmetric eigenstates
\begin{equation}
\begin{split}
\Phi_{\perp ,n}^{S}(z)= B &\Theta ( |z| -a)
\cos(k_{n}a)e^{\kappa _{n}(a-|z|)}  \\ 
+ B&\Theta (a-\left\vert z\right\vert )\cos(k_{n}z)
\end{split}
\end{equation}
and for the antisymmetric eigenstates
\begin{equation}
\begin{split}
\Phi_{\perp ,n}^{A}(z)=C&\Theta (|z| -a)
\text{sgn}\left(z\right) \sin(k_{n}a)e^{\kappa _{n}(a-| z|)} \\ 
+ C&\Theta (a-|z|)\sin(k_{n}z)
\end{split}
\end{equation}
Here $B$ and $C$ are normalization constants and we have defined
\begin{align}
\kappa_{n} &=\frac{\sqrt{2m|E_{\perp ,n}|}}{\hbar} \\
k_{n} &=\frac{\sqrt{2m\big( |\tilde{V}_{\perp }|-|E_{\perp ,n}|\big) }}{\hbar} 
\end{align}%
The eigenvalues $E_{\perp ,n}$ can be determined by the intersection $s_{n}=k_{n}a$ of the curves
\begin{align}
f(ka) &=\tan(ka) \\
g^{S}(ka) &=\frac{\sqrt{(k_{0}a)^{2}-(ka)^{2}}}{(ka)} 
\end{align}
or
\begin{equation}
g^{A}(ka)=\frac{-(ka)}{\sqrt{(k_{0}a)^{2}-(ka)^{2}}}
\end{equation}%
where $k_{0}=\sqrt{2m | \tilde{V}_{\perp}| / \hbar}$. We obtain the eigenvalues
\begin{equation}
E_{\perp ,n}=\frac{\hbar ^{2}s_{n}^{2}}{2ma^{2}}- |\tilde{V}_{\perp}|
\end{equation}

For the Schroedinger equation in the in-plane direction we have
\begin{equation}
\begin{split}
\Big[ -\frac{\hbar ^{2}}{2m}\big[ \frac{1}{r}\frac{\partial }{\partial r}
&\big( r\frac{\partial }{\partial r}\big) +\frac{1}{r^{2}}\frac{\partial
^{2}}{\partial \varphi ^{2}}\big] \\ 
&+ \tilde{V}_{\parallel}\Theta(b-r)\Big]
\Phi _{\parallel }(r,\varphi )=E_{\parallel }\Phi _{\parallel }(r,\varphi )
\end{split}
\label{2DSchroedinger_cylinderQD}
\end{equation}
The effective potential $\tilde{V}_{\parallel }=-V_0+T_{\perp}$
again includes a
contribution from the kinetic energy in growth-direction, which depends on the solution of the eigenvalue problem in $z$ direction, $T_\perp=\tilde{V}_{\perp}-E_{\perp}$. We start the iteration
by setting $\tilde{V}_{\parallel }$ equal to $-V_0$. Because of the symmetry of the potential
around the growth direction, the Hamiltonian commutes with the components of
the angular momentum operator ($\left[ H,l_{z}\right] =0$). Therefore the
two dimensional Schroedinger equation for the angular momentum projection
quantum number $m_{l}$ reduces to an effective one dimensional Schroedinger
equation. Resorting the terms we obtain
\begin{equation}
\begin{split}
\Big[ -\frac{\hbar ^{2}}{2m}\big( \frac{1}{r}\frac{\partial }{\partial r}
&+
\frac{\partial ^{2}}{\partial r^{2}}\big) \\
&+ \tilde{V}_\|\Theta(b-r)+\frac{\hbar ^{2}}{2m}\frac{m_{l}^{2}}{r^{2}}
\Big] \tilde{\Phi}_{\parallel }(r)  =E_{\parallel }\tilde{\Phi}(r)
\end{split}
\end{equation}
where
\begin{equation}
\Phi _{\parallel}(r,\varphi)=\frac{1}{\sqrt{2\pi }}e^{im\varphi }\tilde{\Phi }%
_{\parallel ,m}(\varphi)
\end{equation}
This equation can be cast into the form of a Bessel differential equation. 
A solution of this differential equation inside the
QD is the Bessel function in $m$-th order of the first kind $J_{m}(kr)$. Therefore we obtain inside the QD
\begin{equation}
\widetilde{\Phi }_{\parallel ,m}(r) = A J_{m}(kr)
\end{equation}%
A solution outside the QD is the modified Bessel function $K_{m}(\kappa_{r}r)$. Therefore we obtain outside the QD
\begin{equation}
\widetilde{\Phi}_{\parallel ,m}\left( r\right) =BK_{m}(\kappa _{r}r)
\end{equation}%
At $r=b$, the wave function $\Psi'_{\parallel }$ and $\Psi
_{\parallel }$ have to be continuous. With $k_{0}^{2}=\frac{2m}{\hbar^2}
( -\tilde{V}_{\parallel }) $ and $\kappa_{\mathrm{r}}=\sqrt{k_{0}^{2}-k^{2}}$,
the continuity condition yields
\begin{equation}
N(k)=\frac{J_{m}^{\prime }(kR)}{J_{m}(kR)}-\frac{K_{m}^{\prime }
(\sqrt{k_{0}^{2}-k^{2}}R)}{K_{m}(\sqrt{k_{0}^{2}-k^{2}}R)}=0
\end{equation}
All $k_{n}$ between $0$ and $k_{0}$ with $N(k)=0$ are allowed. For the
eigenvalues of the two dimensional problem we obtain
\begin{equation}
E_{\parallel ,n}=\frac{\hbar^2 \left( k_{n}^{2}-k_{0}^{2}\right) }{2m}
\end{equation}
In summary we have energy levels $E_{\parallel,n,m}$ and wave functions 
$\Phi_{\parallel ,n,m}$ with the quantum numbers $n$ and $m$. The states with
different $m$ and the same $n$ are degenerate.

For the approximate solution of the three dimensional problem we have
to solve the one- and two-dimensional eigenvalues in a self-consistent fashion by determining the updated potentials for the next iteration
step from the eigen-energies of the previous iteration. The procedure is quite efficient, and one obtains converged eigenvalues $E_{n_{z}n_{r}m}$ and wave functions $\Phi _{n_{z}n_{r}m}$ for the
pillbox-shaped QD after only a few iteration steps. The resulting energies and wave functions, obtained using optimized effective parameters, have been checked against
k$\cdot $p-calculations,~\cite{hackenbuchner,homepage} which include strain and piezoelectric effects.

\subsection{Electronic structure of a QD molecule \label{app_molecule}}

After introducing the pillbox model for the electronic structure of 
QDs, we now couple these QDs to molecules. For this purpose we
introduce an ansatz similar to the linear combination of atomic orbitals.

For a detailed description of the calculation we assume a QD
molecule consisting of two QDs, labeled $1$ and $2$. For QD $1$ and $2$ we
assume $N$ and $M$ bound states respectively. Further, for the uncoupled
QDs, we label the wave functions $\Phi _{1}^{n}$ and $\Phi _{2}^{m}$ , the
eigenvalues $\varepsilon _{1}^{n}$ and $\varepsilon _{2}^{m}$ and the
potential $V_{a}$ and $V_{b}$, respectively. To determine the envelope wave
functions $\Phi $, and the corresponding eigenvalues $E$, of the electronically coupled
QDs we use a superposition of the following form 
\begin{equation}
\Phi =\sum_{n}c_{1}^{n}\Phi_{1}^{n}+\sum_{m}c_{2}^{m}\Phi_{2}^{m}
\label{QDM1}
\end{equation}
With the Hamiltonian 
\begin{equation}
\big( H_{0}+V_{a}+V_{b}\big) \Phi =E\Phi  
\label{QDM2}
\end{equation}%
we can apply a multiplication of $( \Phi _{1}^{j})^*$ and
a multiplication of $(\Phi _{2}^{k})^*$ respectively.
Therefore we obtain in matrix notation
\begin{equation}
\begin{pmatrix}
M_{1}^{jn} & M_{2}^{jm} \\ 
M_{3}^{kn} & M_{4}^{km}%
\end{pmatrix}%
\binom{c_{1}^{n}}{c_{2}^{m}}=%
\begin{pmatrix}
A_{1}^{jn} & A_{2}^{jm} \\ 
A_{3}^{kn} & A_{4}^{km}
\end{pmatrix}%
E\binom{c_{1}^{n}}{c_{2}^{m}}  \label{QDM3}
\end{equation}%
where%
\begin{align}
M_{1}^{jn} &=\varepsilon _{1}^{n}\delta _{jn}+\langle \Phi
_{1}^{j}\vert V_{b}\vert \Phi _{1}^{n}\rangle  \label{QDM4}
\\
M_{2}^{jm} &=\varepsilon _{2}^{m}\langle \Phi
_{1}^{j}\vert \Phi _{2}^{m}\rangle +\langle \Phi
_{1}^{j}\vert V_{a}\vert \Phi _{2}^{m}\rangle  \\
M_{3}^{kn} &=\varepsilon _{1}^{n}\langle  \Phi
_{2}^{k}\vert \Phi _{1}^{n}\rangle +\langle \Phi
_{2}^{k}\vert V_{b}\vert \Phi _{1}^{n}\rangle  \\
M_{4}^{km} &=\varepsilon _{2}^{m}\delta _{km}+\langle \Phi
_{2}^{k}\vert V_{a}\vert \Phi _{2}^{m}\rangle
\end{align}%
and
$A_{1}^{jn} =\delta _{jn}$, 
$A_{2}^{jm} =\langle \Phi _{1}^{j}\vert \Phi_{2}^{m}\rangle$,
$A_{3}^{kn} =\langle\Phi _{2}^{k}\vert \Phi_{1}^{n}\rangle$, as well as
$A_{4}^{km} =\delta _{km} $.
This generalized eigenvalue problem can be solved numerically with an eigenvalue-solver.~\cite{LARPACK} Because in this case matrix~$A$ is invertible,
its possible to reduce the generalized eigenvalue problem to an (ordinary)
eigenvalue problem. Therefore we have to solve
\begin{equation}
\left[ 
\begin{pmatrix}
A_{1}^{jn} & A_{2}^{jm} \\ 
A_{3}^{kn} & A_{4}^{km}%
\end{pmatrix}%
^{-1}%
\begin{pmatrix}
M_{1}^{jn} & M_{2}^{jm} \\ 
M_{3}^{kn} & M_{4}^{km}%
\end{pmatrix}%
\right] \binom{c_{1}^{n}}{c_{2}^{m}}=E\binom{c_{1}^{n}}{c_{2}^{m}}
\label{QDM6}
\end{equation}
The eigenvalues and eigenfunctions of this equation have to be
understood as the single-particle result for the electronic structure of the
QD molecule, which can then be used as input in the many-particle
semiconductor Bloch equations.

Furthermore we want to consider an sufficiently weak external electric
field, i.e., an electric field that can be included in the LCAO calculation of the QD molecules.
For electrons, one includes in the potential $V_{a}+V_{b}$ in~\eqref{QDM2} a contribution from the electric field $Fz$ where $F$ is the electric field. For holes, the sign of the electric potential is reversed.
The results of this semi-analytical approach for QD molecules without electric
field were again checked against k$\cdot $p-calculation.~\cite{hackenbuchner,homepage} 
The approach was found to yield a qualitatively correct description of the electronic structure of the QD molecules studied in this paper.

\section{Correlation contributions due to carrier-phonon
and carrier-carrier interaction \label{app_corcontr}}

We investigate a QD/QD
molecule of the ensemble with $M^{e}$ electron and $M^{h}$ hole states
embedded in a quantum well. We make a single-band approximation for the
quantum well and assume that all electron and hole states are spin or
pseudo-spin degenerate, respectively. So every state in the QD/QD molecule
can be labeled by $\lambda =( b,\vec{k}=m,s) $ where $b\in
\{ c,v\} $ is the band index, $m\in \{ 1,\ldots,M^{e}\} /$ $\{ 1,\ldots ,M^{h}\} ,s\in \{ \uparrow,\downarrow\} $. States in the quantum well are labeled by $\lambda=( b,\vec{k}=\vec{k}_{\parallel },s) $. Thus we introduce the notation $\lambda _{1}$ with $\lambda _{1}=( b_{1},\vec{k}_{1},s) $ for all states. With this unified index $\lambda_{1}=( b_{1},\vec{k}_{1},s_{1}) $ a simplification of the
carrier-phonon interaction matrix-elements
\begin{equation}
M_{\lambda _{3}\lambda _{4}}^{\lambda _{1}\lambda _{2}} =M_{\lambda_{3}\lambda _{4}}^{\lambda _{1}\lambda _{2}}\delta_{b_{1}b_{2}}\delta _{b_{3}b_{4}}  \label{CC3}
\end{equation}
and the carrier-carrier interaction
matrix-elements 
\begin{equation}
W_{\lambda _{3}\lambda _{4}}^{\lambda _{1}\lambda _{2}} =W_{\lambda_{3}\lambda _{4}}^{\lambda _{1}\lambda _{2}}\delta_{b_{1}b_{2}}\delta _{b_{3}b_{4}}  \label{CC4}
\end{equation}
follows. 

\begin{widetext}
With the derivation described in section \ref{sec:corcontr} and a generalized notation $\rho _{\lambda _{1}\lambda_{2}}$ for the density matrix
we obtain for dephasing and scattering
processes due to the carrier-phonon interaction
in Markov approximation the following set of equations 
\begin{equation}
S^{cp}_{\lambda _{1}\lambda _{2}} =\frac{\pi }{\hbar }\sum_{\lambda _{3}}\rho_{\lambda _{3}\lambda _{2}}( t) K_{1}^{cp} 
+\frac{\pi }{\hbar }\sum_{\lambda _{3}}\rho _{\lambda _{1}\lambda _{3}}(t)  K_{2}^{cp}  
-\frac{\pi }{\hbar }\sum_{\lambda _{3}}\rho _{\lambda _{3}\lambda_{2}}^*(t) K_{3}^{cp}  -\frac{\pi }{\hbar }
\sum_{\lambda _{3}}\rho _{\lambda _{1}\lambda _{3}}^*( t)  K_{4}^{cp}  
\label{CC11}
\end{equation}
where
\begin{align}
K_{1}^{cp} =&\sum_{\lambda _{4}\lambda _{5}}M_{\lambda _{4}\lambda_{3}}^{\lambda _{1}\lambda _{5}}\rho _{\lambda _{4}\lambda_{5}}^{\ast }( t) 
\sum_{\pm }\left( N+\frac{1}{2}\mp \frac{1}{2}\right) g\left( -\widetilde{\varepsilon }_{\lambda_{5}}+\widetilde{\varepsilon }_{\lambda _{3}}^{\ast }\pm \hbar \omega _{LO}\right) \\
K_{2}^{cp} =&\sum_{\lambda _{4}\lambda _{5}}M_{\lambda _{4}\lambda_{2}}^{\lambda _{3}\lambda _{5}}\rho _{\lambda _{4}\lambda_{5}}^{\ast }(t)  
\sum_{\pm }\left( N+\frac{1}{2}\mp \frac{1}{2}\right) g\left( -\widetilde{\varepsilon }_{\lambda _{3}}+\widetilde{\varepsilon }_{\lambda _{4}}^{\ast }\mp \hbar \omega _{LO}\right)\\
K_{3}^{cp} =&\sum_{\lambda _{4}\lambda _{5}}M_{\lambda _{4}\lambda_{3}}^{\lambda _{1}\lambda _{5}}\rho _{\lambda _{4}\lambda_{5}}\left( t\right)  
 \sum_{\pm }\left( N+\frac{1}{2}\mp \frac{1}{2}\right) g\left( -\widetilde{\varepsilon }_{\lambda _{5}}+\widetilde{\varepsilon }_{\lambda _{3}}^{\ast }\mp \hbar \omega _{LO}\right) \\
K_{4}^{cp} =&\sum_{\lambda _{4}\lambda _{5}}M_{\lambda _{4}\lambda_{2}}^{\lambda _{3}\lambda _{5}}\rho _{\lambda _{4}\lambda_{5}}\left( t\right)  
 \sum_{\pm }\left( N+\frac{1}{2}\mp \frac{1}{2}\right) g\left( -\widetilde{\varepsilon }_{\lambda _{3}}+\widetilde{\varepsilon }_{\lambda _{4}}^{\ast }\pm \hbar \omega _{LO}\right)
\end{align}
Here, we have used the abbreviation $g( z) =\frac{i}{\pi z}$,  and, as
discussed in connection with equation \eqref{complexenergy}, 
$\widetilde{\varepsilon }_{\lambda _{1}}=\varepsilon _{\lambda_{1}}+\Delta \varepsilon -i\Gamma $  is a complex
single-particle energy with an energy shift $\Delta \varepsilon$ and a damping $\Gamma $, which represents
an energetic broadening and thus a finite quasi-particle lifetime due to the electron-phonon interaction.

Further, we need an expression for dephasing and scattering processes due
to carrier-carrier interaction as described in section \ref{sec:corcontr}. 
For the carrier-carrier interaction in Markov approximation one obtains
\begin{equation}
S^{cc}_{\lambda _{1}\lambda _{2}} =\frac{\pi }{\hbar }\sum_{\lambda _{3}\ldots\lambda_{9}}\Big( \rho_{\lambda _{3}\lambda _{2}}( t) K_{1}^{cc} 
+\rho _{\lambda _{1}\lambda _{3}}(t)K_{2}^{cc}  \\
-\rho _{\lambda _{3}\lambda_{2}}^{\ast }(t) K_{3}^{cc}  -\rho _{\lambda _{1}\lambda _{3}}^{\ast }(t) K_{4}^{cc}\Big)
\label{CC12}
\end{equation}
where
\begin{align}
\begin{split}
K_{1}^{cc} &=W_{\lambda _{1}\lambda _{4}}^{\lambda _{6}\lambda _{7}}\left(
W_{\lambda _{3}\lambda _{5}}^{\lambda _{9}\lambda _{8}}\right) ^{\ast }\rho _{\lambda _{4}\lambda _{5}}^{\ast }\left( t\right) \rho _{\lambda_{9}\lambda _{6}}( t) \rho_{\lambda_{7}\lambda _{8}}^{\ast}\left( t\right)  
g\left( -\widetilde{\varepsilon }_{\lambda _{5}}+\widetilde{\varepsilon }_{\lambda _{9}}^{\ast }-\widetilde{\varepsilon }_{\lambda _{8}}+\widetilde{\varepsilon }_{\lambda _{3}}^{\ast }\right) \\
&-W_{\lambda _{1}\lambda _{4}}^{\lambda _{6}\lambda _{7}}\left( W_{\lambda_{3}\lambda _{5}}^{\lambda _{9}\lambda _{8}}\right) ^{\ast }\rho_{\lambda _{4}\lambda _{8}}^{\ast }(t) \rho _{\lambda_{9}\lambda _{6}}\left( t\right) \rho _{\lambda _{7}\lambda _{5}}^{\ast}\left( t\right) g\left( -\widetilde{\varepsilon }_{\lambda _{5}}+\widetilde{\varepsilon }_{\lambda _{9}}^{\ast }-\widetilde{\varepsilon }_{\lambda _{8}}+\widetilde{\varepsilon }_{\lambda _{3}}^{\ast }\right)
\end{split} \\
\begin{split}
K_{2}^{cc} &=W_{\lambda _{3}\lambda _{4}}^{\lambda _{6}\lambda _{7}}\left(
W_{\lambda _{2}\lambda _{5}}^{\lambda _{9}\lambda _{8}}\right) ^{\ast }\rho _{\lambda _{4}\lambda _{5}}^{\ast }\left( t\right) \rho _{\lambda_{9}\lambda _{6}}\left( t\right) \rho _{\lambda _{7}\lambda _{8}}^{\ast}(t) g\left( -\widetilde{\varepsilon }_{\lambda _{3}}+\widetilde{\varepsilon }_{\lambda _{4}}^{\ast }-\widetilde{\varepsilon }_{\lambda _{6}}+\widetilde{\varepsilon }_{\lambda _{7}}^{\ast }\right) \\
&-W_{\lambda _{3}\lambda _{4}}^{\lambda _{6}\lambda _{7}}\left( W_{\lambda_{2}\lambda _{5}}^{\lambda _{9}\lambda _{8}}\right) ^{\ast }\rho_{\lambda _{4}\lambda _{8}}^{\ast }( t) \rho _{\lambda_{9}\lambda _{6}}\left( t\right) \rho _{\lambda _{7}\lambda _{5}}^{\ast}\left( t\right) g\left( -\widetilde{\varepsilon }_{\lambda _{3}}+\widetilde{\varepsilon }_{\lambda _{4}}^{\ast }-\widetilde{\varepsilon }_{\lambda _{6}}+\widetilde{\varepsilon }_{\lambda _{7}}^{\ast }\right) 
 \end{split} 
\end{align}
\begin{align}
\begin{split}
K_{3}^{cc} &=W_{\lambda _{1}\lambda _{4}}^{\lambda _{6}\lambda _{7}}\left(W_{\lambda _{3}\lambda _{5}}^{\lambda _{9}\lambda _{8}}\right) ^{\ast }\text{}\rho _{\lambda _{4}\lambda _{5}}\left( t\right) \rho_{\lambda _{9}\lambda_{6}}^{\ast }\left( t\right) \rho _{\lambda _{7}\lambda _{8}}(t) g\left( -\widetilde{\varepsilon }_{\lambda _{5}}+\widetilde{\varepsilon }_{\lambda _{9}}^{\ast }-\widetilde{\varepsilon }_{\lambda _{8}}+\widetilde{\varepsilon }_{\lambda _{3}}^{\ast }\right) \\
&-W_{\lambda _{1}\lambda _{4}}^{\lambda _{6}\lambda _{7}}\left( W_{\lambda_{3}\lambda _{5}}^{\lambda _{9}\lambda _{8}}\right) ^{\ast }\rho_{\lambda _{4}\lambda _{8}}\left( t\right) \rho _{\lambda _{9}\lambda_{6}}^{\ast }\left( t\right) \rho _{\lambda _{7}\lambda _{5}}( t)
  g\left( -\widetilde{\varepsilon }_{\lambda _{5}}+\widetilde{%
\varepsilon }_{\lambda _{9}}^{\ast }-\widetilde{\varepsilon }_{\lambda _{8}}+\widetilde{\varepsilon }_{\lambda _{3}}^{\ast }\right) 
\end{split} \\
\begin{split}
K_{4}^{cc} &=W_{\lambda _{3}\lambda _{4}}^{\lambda _{6}\lambda _{7}}\left(W_{\lambda _{2}\lambda _{5}}^{\lambda _{9}\lambda _{8}}\right) ^{\ast }\rho _{\lambda _{4}\lambda _{5}}\left( t\right) \rho _{\lambda_{9}\lambda_{6}}^{\ast }\left( t\right) \rho _{\lambda _{7}\lambda _{8}}\left( t\right)  g\left( -\widetilde{\varepsilon }_{\lambda _{3}}+\widetilde{\varepsilon }_{\lambda _{4}}^{\ast }-\widetilde{\varepsilon }_{\lambda _{6}}+\widetilde{\varepsilon }_{\lambda _{7}}^{\ast }\right) \\
&-W_{\lambda _{3}\lambda _{4}}^{\lambda _{6}\lambda _{7}}\left( W_{\lambda_{2}\lambda _{5}}^{\lambda _{9}\lambda _{8}}\right) ^{\ast }\rho_{\lambda _{4}\lambda _{8}}\left( t\right) \rho _{\lambda _{9}\lambda_{6}}^{\ast }\left( t\right) \rho _{\lambda _{7}\lambda _{5}}\left( t\right) g\left( -\widetilde{\varepsilon }_{\lambda _{3}}+\widetilde{\varepsilon }_{\lambda _{4}}^{\ast }-\widetilde{\varepsilon }_{\lambda _{6}}+\widetilde{\varepsilon }_{\lambda _{7}}^{\ast }\right)
\end{split}
\end{align}
\end{widetext}

\end{document}